\documentclass[aps,superscriptaddress,showpacs,nofootinbib,eqsecnum,prd,notitlepage]{revtex4}

\usepackage{graphics} 
\usepackage{epsf} 
\usepackage{eucal}
\usepackage{amssymb} 
\usepackage{amsmath} 
\usepackage{graphicx}
\usepackage{bm} 
\usepackage{dcolumn} 
\usepackage{epsfig}
\usepackage{multirow} 
\usepackage{hyperref} 
\usepackage{subfigure}
\usepackage{flafter}
\usepackage[utf8]{inputenc}

\begin{document}

\title{Suppressing vacuum fluctuations with vortex excitations}

\author{J. F. de Medeiros Neto} \email{jfmn@ufpa.br} 
\affiliation{Faculdade de F\'{\i}sica, Universidade Federal do Par\'a, 
  66075-110, Bel\'em, PA, Brazil}

\author{Rudnei O. Ramos}\email{rudnei@uerj.br}
\affiliation{Departamento de F\'{\i}sica Te\'orica, Universidade do Estado do 
Rio de Janeiro, 20550-013, Rio de Janeiro, RJ,  Brazil}

\author{Carlos Rafael M. Santos} \email{carlos.santos@icen.ufpa.br}
\affiliation{Faculdade de F\'{\i}sica, Universidade Federal do Par\'a,
  66075-110, Bel\'em, PA,  Brazil}

\author{Rodrigo F. Ozela} \email{ozelarf@gmail.com}
\affiliation{Faculdade de F\'{\i}sica, Universidade Federal do Par\'a,
  66075-110, Bel\'em, PA,  Brazil}

\author{Gabriel C. Magalhães} \email{gconduru@ufpa.br}
\affiliation{Faculdade de F\'{\i}sica, Universidade Federal do Par\'a,
  66075-110, Bel\'em, PA,  Brazil}

\author{Van Sérgio Alves} \email{vansergi@ufpa.br}
\affiliation{Faculdade de F\'{\i}sica, Universidade Federal do Par\'a,
  66075-110, Bel\'em, PA,  Brazil}

\begin{abstract}
  
The Casimir force for a planar gauge model is studied considering
perfect conducting and perfect magnetically permeable boundaries. By
using an effective model describing planar vortex excitations, we
determine the effect these can have on the Casimir force between
parallel lines.  Two different mappings between models are considered 
for the system under study, where generic boundary conditions can be 
more easily applied and the Casimir force be derived in a more 
straightforward way. It is shown that vortex excitations can be an 
efficient suppressor of vacuum fluctuations. In particular, for the model
studied here, a planar Chern-Simons type of model that allows for the 
presence of vortex matter, the Casimir force is found to be independent of 
the choice of boundary conditions, at least for the more common types, like
Neumann, perfect conducting and magnetically permeable boundary conditions. 
We give an interpretation for these results and some possible applications 
for them are also discussed.

\end{abstract}


\maketitle

\section{Introduction}
\label{intro}

There has been considerable interest in studying the validity of
Newton's  gravitational law at sub-millimeter scales and well below
that (for a recent review see, e.g., Ref.~\cite{Will:2014xja}).  There
is a possibility that with these experiments deviations from  the
standard power-law behavior could be found, thus, possibly  probing
phenomena like modified gravity scenarios predicted by  string theory,
or by physics beyond the standard model of particle  physics. {}For
example, compactified extra spatial dimensions in  string theory could
lead to a modification of the quadratic  power-law behavior, depending on the
number of extra dimensions. Also, physics beyond the standard  model
of particle physics can produce  Yukawa-type corrections for the
gravitational force (for a  comprehensive review, see also,  e.g.,
Ref.~\cite{Adelberger:2003zx}  and references therein).

Laboratory experiments measuring gravity related forces at  extremely
small scales pose some extraordinary challenges. One of these
challenges for probing forces  at such very small scales is to
distinguish gravitation-like interactions from other effects that can come
from quantum phenomena, most notably the Casimir  force~\cite{casimir}, 
which can potentially dominate gravity effects
by  several orders of magnitude at distances of the order of the
micrometer  and below that.  In fact, the fast recent developments on
laboratory experiments measuring the Casimir
force~\cite{review,Klimchitskaya:2009cw} have also helped to put some
strong constraints on the level of possible corrections to
gravity~\cite{Bezerra:2010pq}.  On the other hand, it is also highly
desirable to devise ways of either isolating the Casimir effect, or to
suppress it up to the level of precision that can be found in those
experiments. Recently, graphene~\cite{graphene} has been proposed for
such purpose due to its  extraordinary absorption properties,  which
could effectively function as a shield for quantum vacuum
fluctuations.  It is also important to look for other types of
materials that can be as versatile in terms of been easily produced
and also with tunable properties under laboratory conditions. One such
possibility could be, for example, the use of superconducting films. 

It is known that superconducting films can have magnetic vortex
excitations. Most of the properties of these systems can be described
in terms of  planar gauge systems. We recall that planar gauge field
theories, in particular Chern-Simons (CS) type of models, have  long
been recognized as important for understanding several physical
phenomena that can be well approximated as planar ones, like
high-temperature  superconductivity and the fractional quantum Hall
effect, just to cite a few  examples (see, e.g., Ref.~\cite{reviewCS}
and references therein).  The Casimir force in the presence of
condensed vortices in a plane was studied previously in
Ref.~\cite{felipeRudnei} from the point of view  of the
particle-vortex duality, where an effective description of vortex
excitations was made in terms of a Maxwell-Proca-Chern-Simons (MPCS)
model. 

The study of the Casimir force in the presence of vortex excitations
carried out in Ref.~\cite{felipeRudnei} was based on a particular
mapping  existing between the MPCS model and a model of two
noninteracting massive scalar fields. Since the Casimir force is well
known for the latter case, the corresponding result for the former
could be easily determined. This mapping, however, severely restricted
the form of the boundary conditions (BC) considered there. In
particular, the connection between the different model Hamiltonians
was only possible in the case of Neumann BC for the scalar field and,
in this sense, the form of the mathematical transformations has
implied in the consideration of a specific type of BC for the scalar
and vector fields.  Also, the connection was only possible for the
simplest geometry treated there to compute the Casimir force, i.e.,
the force between parallel lines,  and could not be generalized to
other geometries.

It is a much desirable solution to explore appropriate mappings
between models that can be used in the determination of the Casimir
effect, where the above mentioned restrictions can be avoided.  
In this paper, we will consider two known relationships associated
with the MPCS model: (a) the connection of the MPCS model with a sum of
a self-dual and an anti-self dual  Proca-Chern-Simons (PCS) 
models~\cite{BanerjeeKumar2}, and (b) the connection of the MPCS model with
a sum of two Maxwell-Chern-Simons (MCS) models~\cite{Banerjee:1998xv,BanerjeeKumar2}.
The advantage of both associations, as compared to that related to
scalar degrees of freedom~\cite{felipeRudnei,BJP}, is that a direct relation between 
the original and final fields can be made very clear. This in turn facilitates 
the connection between the BC and also the calculation of the Casimir force.

As we are going to see in this paper, the difficulties met with the original 
mapping used in Ref.~\cite{felipeRudnei} are removed.
In the case (a) listed above, we make use of the intrinsic properties of
self duality and anti-self duality of the PCS models. This allows us 
to define mathematically the BC in terms of the Green's functions. 
Then, performing the calculations again in the case (b) listed above,
help to confirm our results.
As we are going to show, the use of the relation (a) facilitates our 
calculations, because we can make use of the symmetry of the resulting 
PCS models and the final form of the energy-momentum tensor. Another important
benefit provided by the relation (a) is that the number of differential
equations that we need to solve and the number of required Green's functions
are smaller, when compared to the case (b), as we are going to see.

Our objectives in this work are twofold. {}Firstly, by using more
general mappings than the one used in Ref.~\cite{felipeRudnei}, we can compute the 
Casimir force in the cases of more realistic and physically relevant BC and
geometries. Secondly, with the use of a different BC, we can determine
any possible effect that might have on the Casimir force. 
We shall derive results for two BC of interest, i.e., 
for perfect conducting and for perfect magnetically 
permeable boundaries. We also consider another type of (Neumann) BC,
previously considered in Ref.~\cite{felipeRudnei}, and confirm 
the result found there. We shall still use for convenience and simplicity 
the simplest geometry of parallel lines, but our results can be 
extended to other more complex geometries,  which we leave for a future work.
It is explicitly shown that, for the model
studied here, the Casimir force is found to be independent of the choice of BC used.

Given the many approximations and considerations assumed in our
calculation (which will be discussed below in the text), the use of
the results that we have obtained in this work to the high precision
gravity and Casimir experiments that were mentioned above may sound
too optimistic and, thus, should not be taken literally in that
context. However, the present results point to effects that can be of
relevance in the future planning of these experiments. Nevertheless,
the present work  is of theoretical interest, where some novel aspects
related to topological (vortex) excitations are considered, along also
with issues regarding the use of different BC in the computation of
the Casimir effect. 

The remainder of this work is organized as follows. In
Sec.~\ref{sec2}, we summarize the connection of the MPCS model as an
effective vortex-particle dual to the Chern-Simons-Higgs (CSH) model. We also summarize
the mathematical relations that connect the MPCS model in terms of
a self-dual and an anti-self dual PCS models and also in terms of
a sum of two MCS models. In Sec.~\ref{sec3}, we analyze the relation between
the original vector field of the MPCS model and the new fields
associated with the two PCS models and give the relevant equations
needed to evaluate the Casimir force. This evaluation is done considering  
the cases of perfect conductor and also perfect magnetically permeable lines 
at the boundaries.
In Sec.~\ref{sec4}, we check and confirm our results to be independent
of the mapping used, by considering this time 
the connection between the MPCS and two MCS
models, re-deriving our results again for both cases of perfect magnetically 
permeable and  perfect conductor boundaries. 
In Sec.~\ref{sec5}, based on the symmetries and constraints of the models
studied, we explain the reason for the independence of the Casimir force on 
the BC considered in the calculations.
In Sec.~\ref{sec6} we analyze and discuss the Casimir force obtained in the context
of a vortex condensate.
{}Finally, in Sec.~\ref{sec7}, we give our concluding remarks and discuss other 
possible applications and implications of the results derived in this work. 

\section{The MPCS model as an effective dual vortex
description and its mapping onto two PCS models}
\label{sec2}

It has been shown in Ref.~\cite{dual} (for earlier derivations, see
for example Ref.~\cite{Wen:1988uw})  that vortex excitations in a
CSH model can be expressed effectively in terms
of a dual equivalent theory (for
applications of similar duality ideas in planar systems of interest in
condensed matter and that makes use of the particle-vortex duality in
Chern-Simons type of models, see Ref.~\cite{Burgess:2000kj} and
references therein). This effective model for vortices, in turn, can
be expressed in the form of a MPCS model, when both the scalar Higgs
field and  the vortex field are in their symmetry broken vacuum
states, $\rho_0 \neq 0$ and $\psi_0\neq 0$, respectively. The Lagrangian
density of the MPCS model can be expressed as~\cite{felipeRudnei}:

\begin{equation}
\label{LMPCS}
\mathcal{L}= - \frac{1}{4} F^{\alpha \beta} F_{\alpha \beta} +
\frac{m^{2}}{2}A^{\alpha}A_{\alpha} + \frac{\mu}{4}\epsilon^{\alpha
  \beta \lambda} A_{\alpha}\partial_{\beta}A_{\lambda}~,
\end{equation}
where 

\begin{eqnarray}
&& m \equiv 4\pi \rho_0 \psi_0, \label{massm}\\  &&\mu \equiv  2 e^2
  \rho_0^2/\Theta,
\label{massmu}
\end{eqnarray}
and $\Theta$ is the Chern-Simons parameter of the original CSH model,
from which Eq.~(\ref{LMPCS}) is derived.  

In addition to the connection of the above model with a dual vortex equivalent one, the
MPCS model given by Eq.~(\ref{LMPCS}) can be mapped in
a sum of a self dual and an anti-self dual PCS 
models~\cite{BanerjeeKumar2} or, also, in terms of a sum of 
two MCS models~\cite{Banerjee:1998xv,BanerjeeKumar2}.
As we will discuss later on, these
associations will simplify considerably the calculation of the Casimir
force. {}For completeness, let us briefly review below these two
considerations concerning the model given by Eq.~(\ref{LMPCS}).


\subsection{The effective dual vortex description for the MPCS model}

Chern-Simons gauge field theories can exhibit many features of
relevance in different contexts. One of these features, which is of
particular importance in our study, is the possibility of having
topological vortex solutions when these models are coupled to symmetry
broken scalar potentials~\cite{khare}.  {}For instance, we can
consider the CSH model described by the Euclidean action

\begin{equation}
S_E[h_\mu, \eta,\eta^*] = \int d^3 x \left[ - i\frac{\Theta}{4}
  {}\epsilon_{\mu \nu \gamma}h_{\mu}{}H_{\nu \gamma} + |D_\mu \eta|^2
  + V(|\eta|) \right]~,
\label{actionE}
\end{equation}
where $H_{\mu \nu} = \partial_\mu h_\nu - \partial_\nu h_\mu$, $D_\mu
\equiv \partial_\mu + ieh_\mu$ and $\eta$ is a  complex scalar field,
with a  non-null vacuum expectation value (VEV) obtained from a
symmetry breaking polynomial potential  $V(|\eta|)$. {}For instance,
for a potential given by $V(|\eta|)=e^4\left( |\eta|^2
-\nu^2\right)^2|\eta|^2/\Theta^2$, the field equations for the model
(\ref{actionE}) has nontrivial vortex solutions given
by~\cite{Jackiw:1990aw}  

\begin{equation}
\eta_{\rm vortex} = \xi(r) \,\exp(i n \chi)~, \ \  \ \ \ \ h_{\mu,
  {\rm vortex}} = \frac{n}{e} h(r)\, \partial_\mu \chi~,
\end{equation}
where $n$ is an integer that represents the vortex charge, while
$\xi(r)$ and $h(r)$  are the (vortex profile) functions obtained from
the solutions of the classical field  differential equations,
subjected to the BC $\lim_{r \to 0}\xi(r) = 0$, $\lim_{r \to
  \infty}\xi(r) = \nu$,  $\lim_{r \to 0}h(r) = 0$ and $\lim_{r \to
  \infty}h(r) = 1$.  The presence of vortex excitations means
that the phase of the scalar field, $\phi = \rho
\exp(i\chi)/\sqrt{2}$,  is a multivalued function. The phase $\chi$
can then in general be expressed in terms of a regular (single valued)
and a singular part as $\chi(x) = \chi_{\rm reg}(x) + \chi_{\rm
  sing}(x)$.  The vortex excitations can be made explicit in the
action by functionally integrating over the regular phase, while
leaving explicitly the dependence of the singular phase in the action.
This procedure can be done by the so-called dual transformations (see,
e.g., Refs.~\cite{dual,Kim-Lee} for a detailed account for this
procedure). The final result can be expressed in terms of a dual
action, written in terms of a complex scalar field $\psi$
(representing  quantized vortex excitations) and a new gauge field
$A_{\mu}$, which is related to the original fields by the relation
$\rho^2(\partial_\mu \chi + e h_\mu) = (\sigma/(2 \pi e))
\epsilon_{\mu \nu \gamma}  \partial_\mu A_\gamma$, where $\sigma$ is
an arbitrary parameter with mass dimension.  The final dual action can be
expressed in the form~\cite{dual}

\begin{equation}
\label{S_ini}
S_{\rm dual} = \int d^3 x  \left[ \frac{\sigma^2}{16 \pi^2 e^2
    \rho_0^2}F_{\mu\nu}^2 +  i \frac{\sigma^2}{8 \pi^2 \Theta}
  \epsilon_{\mu \nu \gamma} A_\mu \partial_\nu A_\gamma +  \left|
  \partial_{\mu}\psi + i\frac{2\sigma}{e} A_{\mu}\psi \right|^2 +
  V_{\rm vortex}(|\psi|) + {\cal L}_{G} \right] \,,
\end{equation}
where $F_{\mu\nu} \equiv \partial_{\mu}A_{\nu} -
\partial_{\nu}A_{\mu}$, $V(|\psi|)$ is the effective potential term
for the vortex field, with a VEV  $\psi_0$, and ${\cal L}_{G}$ is a
gauge fixing term. 

When the system is taken deep inside its vortex condensed phase, 
we can take the London type approximation for the vortex
field~\cite{kleinert},  where $|\psi| \to \psi_0/\sqrt{2}$.  In this
case, we can neglect the derivative of $\psi$ that appears in
Eq.~(\ref{S_ini}).  We can also choose $\sigma \equiv 2\pi e \rho_0$,
so that Eq.~(\ref{S_ini}) can then be finally  rewritten in the form
of the MPCS model with the (Minkowski) Lagrangian density given by
Eq.~(\ref{LMPCS}).

\subsection{Mapping the MPCS model onto two PCS models}
\label{2PCS}

To compute the Casimir force for  the MPCS model, we could in
principle start directly from Eq.~(\ref{LMPCS}) and use standard
methods based on the vacuum expectation values for the space-space and
time-time components of the  energy-momentum tensor (like, e.g., those
discussed in Ref.~\cite{Milton_1990}). This procedure leads, however,
to a hard to solve  system of partial differential equations (PDE). It
turns out that it is much simpler to express the original
model, Eq.~(\ref{LMPCS}), in  terms of an equivalent one that can be
easily treated mathematically.  In particular, we want to have a well
defined mapping between the fields in each model, such that we can
unequivocally establish their behaviors at the physical boundaries of
the system. Such mapping must imply in a direct correspondence between
the BC considered for the MPCS and its equivalent model, resulting in
a one-to-one mapping between the Casimir forces for the models
involved. One of such possibility is to follow the proposal of
Refs.~\cite{Banerjee:1998xv,BanerjeeKumar2}, where the MPCS of
Eq.~(\ref{LMPCS}) is mapped into a doublet consisting of a
self-dual and an anti self-dual PCS models in 2+1 dimensions.  One of
the advantages of this procedure is that a direct relation between the
original and final fields can be made very clear, which facilitates
the connection between the BC. Besides, it also allows the use of
different BC and, eventually, it can also be generalized to different
geometries, as opposite to the case treated originally in
Ref.~\cite{felipeRudnei}.

{}Following in particular Ref.~\cite{BanerjeeKumar2}, we consider a
doublet consisting of an anti self-dual and a self-dual PCS models,
represented, respectively, by the Lagrangian densities,

\begin{equation}
\mathcal{L}_{-}=-\frac{1}{2}\epsilon_{\mu\nu\beta}g^{\mu}\partial^{\nu}
g^{\beta} +  \frac{m_{-}}{2}g_{\mu}g^{\mu},
\label{Lmenos}
\end{equation}     
and

\begin{equation}
\mathcal{L}_{+}=\frac{1}{2}\epsilon_{\mu\nu\beta}f^{\mu}\partial^{\nu}
f^{\beta} +  \frac{m_{+}}{2}f_{\mu}f^{\mu},
\label{Lmais}
\end{equation}
where $f_{\mu}$ and $g_{\mu}$ are two independent vector fields. By
making use of a  soldering field  $W_{\mu}$ with no dynamics, it is a
simple exercise to obtain, from the combination of $\mathcal{L}_{+}$
and $\mathcal{L}_{-}$, a final Lagrangian density that does not depend
on $W_{\mu}$. {}For example, we can define an intermediate Lagrangian
density given by

\begin{equation}
\mathcal{L} = \mathcal{L}_{-}(g) + \mathcal{L}_{+}(f) - W_{\mu}\left[
  J^{\mu}_{-}(g) +  J^{\mu}_{+}(f) \right] + \frac{1}{2}(m_{+} +
m_{-})W_{\mu}W^{\mu},
\label{LInt}
\end{equation}    
where $J^{\mu}_{\pm}$ are defined by 

\begin{eqnarray}
&& J^{\mu}_{+}(f) \equiv \sqrt{m_{+}}f^{\mu} +
  \epsilon^{\mu\alpha\beta}\partial_{\alpha}f_{\beta}, \\  &&
  J^{\mu}_{-}(g) \equiv \sqrt{m_{-}}g^{\mu} -
  \epsilon^{\mu\alpha\beta}\partial_{\alpha}g_{\beta}.
\end{eqnarray}    
In the generating functional associated with (\ref{LInt}), $W^{\mu}$
plays the role of an auxiliary field, which can be eliminated by a
direct integration (another way of seeing the auxiliary role of $W^{\mu}$
is by the use of its equation of motion). The resulting final
Lagrangian density can then be written as

\begin{equation}
\mathcal{L}=-\frac{1}{4}F_{\mu\nu}F^{\mu\nu} +  \frac{( m_{-}-
  m_{+})}{2} \epsilon_{\mu\nu\beta}A^{\mu}\partial^{\nu}A^{\beta} +
\frac{1}{2}m_{+}m_{-}A_{\mu}A^{\mu},
\label{hep1}
\end{equation} 
where $A_{\mu}$ is a new vector field, related to $f_{\mu}$ and
$g_{\mu}$ by

\begin{equation}
A_{\mu} \equiv \frac{1}{\sqrt{m_{+}+m_{-}}}(f_{\mu}-g_{\mu})\,,
\label{Afg}
\end{equation}
and $m_{+}$ and $m_{-}$ are related to the original mass parameters
$\mu$ and $m$ of Eq.~(\ref{LMPCS}) by 

\begin{eqnarray}
m_{-} -  m_{+} & = & \mu/2\,, \label{mMenosmMais_1} \\   
m_{+}m_{-} & = & m^{2}\, .
\label{mMenosmMais_2}
\end{eqnarray}
It is important to note that in Eq.~(\ref{Afg}) we consider 
that $m_{+}$ and $m_{-}$ are both positive. This consideration
implies that $m^{2} > 0$. Thus, Eqs.~(\ref{mMenosmMais_1}) and (\ref{mMenosmMais_2})
imply that 

\begin{equation}
m_{\pm}  =  \mp \frac{\mu}{4} + \sqrt{ \frac{\mu^2}{16} + m^{2}}\,.
\label{mMenosMais}
\end{equation}

The result of the detailed study of the relation between $\mathcal{L}$
and $\mathcal{L}_{+} + \mathcal{L}_{-}$ shows a complete equivalence
between them~\cite{BanerjeeKumar2}, i.e., $\mathcal{L} =
\mathcal{L}_{+} + \mathcal{L}_{-}$.  Hence, it is straightforward to
perceive that the Casimir force related to the original MPCS model can
be written as the sum of the Casimir forces associated with
$\mathcal{L}_{+}$ and $\mathcal{L}_{-}$.  Also, the simple relation
between $f_{\mu}$, $g_{\mu}$ and $A_{\mu}$, given by Eq.~(\ref{Afg}),
implies in a direct determination of the BC considered for  $f_{\mu}$ and $g_{\mu}$, in terms of those 
considered for $A_{\mu}$. We can also
conclude from Eq.~(\ref{Afg}) that, in principle, there is no
restriction for the BC to be considered for  $A_{\mu}$ (which will be
associated with the BC for $f_{\mu}$ and $g_{\mu}$), as long as they
are  mathematically and physically acceptable.  We also note that
determining the Casimir force related to a PCS model is rather simpler
than determining the force  for the MPCS model directly, as we will
discuss in the next section. 

\subsection{The MPCS model written in term of two MCS models}
\label{2MCS}

Alternatively, we can also use the equivalence between 
the MPCS model and a doublet of MCS models, given in Ref.~\cite{BanerjeeKumar2}. 
These two MCS models will be written in terms of two gauge fields $P_\mu$ and $Q_\mu$,
respectively, which can be conveniently rescaled, when compared with their analogues 
considered in Ref.~\cite{BanerjeeKumar2}. We can write the Lagrangian densities for
the two MCS models as 
 
\begin{equation}
\tilde{\mathcal{L}}_{-}(P) = -\frac{1}{4}P_{\mu\nu}P^{\mu\nu}
+\frac{1}{2} m_{-} \epsilon_{\mu\nu\beta}P^{\mu}\partial^{\nu}
P^{\beta}~, 
\label{LmenosP}
\end{equation}     
and

\begin{equation}
\tilde{\mathcal{L}}_{+}(Q) = -\frac{1}{4}Q_{\mu\nu}Q^{\mu\nu}
-\frac{1}{2}  m_{+} \epsilon_{\mu\nu\beta}Q^{\mu}\partial^{\nu}
Q^{\beta}\, ,
\label{LmaisQ}
\end{equation}
where $P^{\mu\nu} = \partial^{\mu}P^{\nu} - \partial^{\nu}P^{\mu}$, 
and $Q^{\mu\nu}= \partial^{\mu}Q^{\nu} - \partial^{\nu}Q^{\mu}$. The masses $m_+$ and $m_-$ in Eqs.~(\ref{LmenosP})
and (\ref{LmaisQ}) are the same as the ones defined in Eq.~(\ref{mMenosMais}).

The two gauge fields $P_\mu$ and 
$Q_\mu$ are connected to the original gauge field $A_\mu$ of the MPCS 
model by

\begin{equation}
\label{APQ}
A_{\mu} \equiv \frac{1}{\sqrt{m_{+}+m_{-}}}(\sqrt{m_{-}}P_{\mu} - \sqrt{m_{+}}Q_{\mu})\,.
\end{equation}

The relation between the doublet of MCS models, Eqs.~(\ref{LmenosP})
and (\ref{LmaisQ}), with the MPCS model (\ref{LMPCS}) is stablished
in a similar fashion as in the case of the previous subsection.
By using this time a tensor field $B_{\mu\nu}$ connecting the two
Lagrangian densities (\ref{LmenosP})
and (\ref{LmaisQ}), we have that

\begin{equation}
\mathcal{L}= \tilde{\mathcal{L}}_{-}(P) + \tilde{\mathcal{L}}_{+}(Q) -
\frac{1}{2} B_{\mu \nu} \left[
  J^{\mu\nu}_{-}(P) +  J^{\mu\nu}_{+}(Q) \right] - \frac{m_+ + m_-}{4 m_{+} 
m_{-}} B_{\mu \nu}B^{\mu\nu},
\label{LInt2}
\end{equation}    
where $J^{\mu}_{\pm}$ are defined by 

\begin{eqnarray}
&& J_{\mu\nu}^{+}(Q) \equiv - Q_{\mu\nu} -
  m_{+}\epsilon_{\mu\nu\beta}Q^{\beta}, 
	\label{FonteQ}
	\\  &&
	\label{FonteP}  
	J_{\mu\nu}^{-}(P) \equiv - P_{\mu\nu} +
  m_{-} \epsilon_{\mu\nu\beta}P^{\beta}.
\end{eqnarray}    
Again, considering the relation between the fields given in Eq.~(\ref{APQ}), we can 
eliminate the auxiliary feld $B_{\mu \nu}$, reproducing once again the original
MPCS model.

It is important to realize that in both mappings described above, the number of degrees 
of freedom is preserved. It is noteworthy to realize that in a MCS model the mass term 
for the gauge field is of topological origin. Each MCS model has only one (transverse) 
polarization degree of freedom. However, in the MPCS model, the explicit mass term for 
the gauge field implies that there are now two polarization degrees of freedom for 
the gauge field. The number of degrees of freedom is preserved in the two mappings used. 
The duality between these different types of gauge models has also been discussed extensively 
in the literature before. {}For example, in Ref.~\cite{Dalmazi:2006yv} this issue is 
discussed in terms of an interpolating master action and how it explains the doubling of fields, 
yet preserving the number of degrees of freedom.

{}Finally, it is important to also note that while the
association of the vortex excitations in the CSH model with the  MPCS model given in  
Eq.~(\ref{LMPCS}) is only valid within the approximations considered
in the previous subsection (e.g., for a special Higgs  potential, no
vortex interactions,  and the use of a London-type limit for the
Higgs and vortex fields), the relation between the MPCS and 
PCS models is exact. The same can be said with respect to the MCS models.

\section{The Casimir force for the MPCS model expressed in terms 
of a doublet of PCS models}
\label{sec3}

In this section, we will use an analogous procedure as used, e.g., in
Ref.~\cite{Milton_1990} to calculate the Casimir forces
associated with $\mathcal{L}_{+}$ and $\mathcal{L}_{-}$, given by
Eqs.~(\ref{Lmais}) and  (\ref{Lmenos}), respectively. 

In the
following, we have adopted the notation $X\equiv x^\mu = (t,x,y)$ 
and considered the metric tensor $\eta^{\mu\nu}=
\mbox{diag}(1, - 1, - 1)$. The physical boundaries are placed in  $x = 0$
and $x = a$.

The Casimir force (per unit length) for the MPCS model is determined
from the 11-component of the  energy-momentum tensor,

\begin{equation}
f\equiv (\mbox{force/lenght})_{\mbox{MPCS}}  =  \left\langle\,
{T^{11}}_{MPCS}\, \right\rangle\Bigr|_{x=0\,{\rm and}\, x=a}\, ,
\label{pressaoCasimir}
\end{equation} 
which can also be written, according to the results shown in the
previous section,  as 

\begin{equation}
f = \left[\left\langle\,  T^{11}_{-}\, \right\rangle 
+  \left\langle\, T^{11}_{+}\, \right\rangle\right]\Bigr|_{x=0\,{\rm and}\, x=a}\,,
\label{pressaoCasimirFG}
\end{equation} 
where $T^{11}_{-}$ is the energy-momentum tensor component obtained
from $\mathcal{L}_{-}$, given by Eq.~(\ref{Lmenos}), while  $T^{11}_{+}$ is the one obtained from
$\mathcal{L}_{+}$, given by Eq.~(\ref{Lmais}). As it is well known, the CS term does not contribute 
to the symmetric energy-momentum tensor, since it is given in terms of 
the derivative of the action with respect to the metric tensor and the 
CS term does not depend on this metric~\cite{Milton_1990,Dunne}. Thus,
we obtain:

\begin{eqnarray}
&&T^{\mu\nu}_{-} =
  -\eta^{\mu\nu}\frac{m_{-}}{2}g_{\alpha}g^{\alpha},
\label{T-} 
\\ &&T^{\mu\nu}_{+} =
-\eta^{\mu\nu}\frac{m_{+}}{2}f_{\alpha}f^{\alpha}\,.
\label{T+}
\end{eqnarray} 

Equation (\ref{pressaoCasimirFG}) can be written in terms of the
Green's functions for the gauge fields $f^\mu$ and $g^\mu$, 
$G^{\mu \nu}_+(X,X')=i\langle {\hat T}[f^\mu(X)f^\nu(X')]\rangle$ and
$G^{\mu  \nu}_-(X,X')=i\langle {\hat T}[g^\mu(X)g^\nu(X')]\rangle$, respectively.  
{}For example, using Eq.~(\ref{T-}), we can write

\begin{equation}
\langle T^{11}_{-}(X) \rangle =  -i\frac{m_{-}}{2}\lim_{X^{\prime}\to
  X}\left[ G_{-}^{00}(X,X^{\prime}) - G_{-}^{11}(X,X^{\prime}) -
  G_{-}^{22}(X,X^{\prime})\right]\,,
\label{tUmUmMenos}
\end{equation}
and similarly for $\langle T^{11}_{+}(X) \rangle$.

The Green's functions for $f^\mu$ and $g^\mu$ can be
derived from the Euler-Lagrange equations for the fields as usual:

\begin{eqnarray}
&& m_{-}g_{\mu}(X) - \epsilon_{\mu\beta\nu}\partial^{\beta}g^{\nu}(X)
  + J_{(-)\mu}(X)=0\,,
\label{eqMotion_g}  
\\ && m_{+}f_{\mu}(X) +
\epsilon_{\mu\beta\nu}\partial^{\beta}f^{\nu}(X) + J_{(+)\mu}(X)=0\,,
\label{eqMotion_f}
\end{eqnarray}
where $ J_{(-)\mu}$ and $ J_{(+)\mu}$ are the source terms.  The
formal solutions to Eqs.~(\ref{eqMotion_g}) and (\ref{eqMotion_f}) are

\begin{eqnarray}
&&g^{\mu}(X) = \int G_{-}^{\mu\alpha}( X,
  X^{\prime})J_{(-)\alpha}(X')d X^{\prime}\,,
\label{gGJ}
\\ &&f^{\mu}(X) = \int G_{+}^{\mu\alpha}( X,
X^{\prime})J_{(+)\alpha}(X')d X^{\prime}\,,
\label{fGJ}
\end{eqnarray}
and

\begin{eqnarray}
&&m_{-}G_{-}^{\mu\alpha} - {\epsilon^{\mu}}_{\beta\nu}
  \partial^{\beta}G_{-}^{\nu\alpha}+\delta(X-X^{\prime})\eta^{\mu\alpha}=0\,,
\label{pdeForGMenos}
\\ &&m_{+}G_{+}^{\mu\alpha} + {\epsilon^{\mu}}_{\beta\nu}
\partial^{\beta}G_{+}^{\nu\alpha}+\delta(X-X^{\prime})\eta^{\mu\alpha}=0\,.
\label{pdeForGMais}
\end{eqnarray}

Note that, unlike the calculations followed in
Refs.~\cite{Milton_1990,DanEdneyFelSil} (where the Green's functions
for the field's duals were used), we work directly
in terms of the Green's functions for the fields themselves ($f_{\mu}$ and $g_{\mu}$). 
This would also be the case if we had decided to work with the MPCS model 
directly. This fact can be seen as a consequence of the fact that the Proca term, $m^2A_{\mu}A^{\mu}/2$, cannot 
be written in terms of the dual of $A_{\mu}$. But if we had decided to work with the MPCS model 
directly (without ``transforming'' it to a doublet of PCS models beforehand as we are
proceeding here), the resulting system of second-order differential equations would be 
more difficult to solve, when compared to the one that we have in the present
case~\cite{Milton_1990,DanEdneyFelSil}. The
transformations taken here simplify the calculations significantly,
since the system of equations with which we have to deal with is
relatively easier to solve, given by Eqs.~(\ref{pdeForGMenos}) and
(\ref{pdeForGMais}). 

Using the Fourier transforms in time and in the transverse coordinate $y$
for $G_{\pm}^{\mu\nu}(X, X^{\prime})$,   

\begin{equation}
G_{\pm}^{\mu\nu}(X, X^{\prime}) =
\int\frac{d\omega}{2\pi} e^{-i\omega(t-t')} 
\int\frac{dk}{2\pi}e^{ik(y-y')} 
\mathcal{G}_{\pm}^{\mu\nu}(k,\omega,x,x^{\prime})\,,
\label{transformadaDeGmenos}
\end{equation}
we can write:

\begin{eqnarray}
\langle  T^{11}_{\pm} \rangle & = &
-i\frac{m_{\pm}}{2}\lim_{X^{\prime}\to
  X}\int\frac{d\omega}{2\pi}e^{-i\omega(t-t')} 
\int\frac{dk}{2\pi}\, e^{ik(y-y')}\times  \nonumber \\   & &
\left[\, \mathcal{G}_{\pm}^{00}(k,\omega,x,x^{\prime}) -
  \mathcal{G}_{\pm}^{11}(k,\omega,x,x^{\prime}) -
  \mathcal{G}_{\pm}^{22}(k,\omega,x,x^{\prime}) \right], 
\label{t11mais_menos}
\end{eqnarray}
and the Casimir force (per unit length) can be expressed as

\begin{eqnarray}
f &=& \langle  {T^{11}}_{MPCS}\rangle \Bigr|_{x=0\,{\rm and}\, x=a}
\nonumber \\ \nonumber \\ & = & \left[\langle T^{11}_{-}(X) \rangle +   \langle T^{11}_{+}(X)\rangle 
\right]\Bigr|_{x=0\,{\rm
    and}\, x=a}     \nonumber \\ \nonumber \\   & = & \!-i\lim_{X^{\prime}
  \rightarrow X} \left\{ \int\frac{d\omega}{2\pi}e^{-i\omega(t-t')}\!
\int\frac{dk}{2\pi}e^{ik(y-y')}\!  \left[
  \frac{m_{-}}{2}\left(\mathcal{G}_{-}^{00} \!-\!
  \mathcal{G}_{-}^{11} \!-\!  \mathcal{G}_{-}^{22}\right) \!+\!
  \frac{m_{+}}{2}\left(\mathcal{G}_{+}^{00} \!-\! \mathcal{G}_{+}^{11}
  \!-\!  \mathcal{G}_{+}^{22}\right)\right] \right\} \Bigr|_{x=0\,{\rm
    and}\, x=a} .
\label{T11_exp1}
\end{eqnarray}

The components $\mathcal{G}_{\pm}^{00}$, $\mathcal{G}_{\pm}^{11}$ and
$\mathcal{G}_{\pm}^{22}$ are obtained from the solutions of the
following systems of PDE (where $x$ stands for $x^1$):

\begin{equation}
\left\{
\begin{array}{ccc}
-ik\mathcal{G}_{-}^{01}+m_{-}\mathcal{G}_{-}^{11}+
i\omega\mathcal{G}_{-}^{21}&=&\delta(x-x^{\prime}),\\ \\     
m_{-}\mathcal{G}_{-}^{01}-ik\mathcal{G}_{-}^{11}+
\partial_x\mathcal{G}_{-}^{21}&=&
0,\\   \\ \partial_x\mathcal{G}_{-}^{01}-
i\omega\mathcal{G}_{-}^{11}+m_{-}\mathcal{G}_{-}^{21}&=& 0,
\label{g1}
\end{array}
\right.
\end{equation}

\begin{equation}
\left\{
\begin{array}{ccc}
-m_{-}\mathcal{G}_{-}^{00} + ik\mathcal{G}_{-}^{10} -
\partial_x\mathcal{G}_{-}^{20}&=&\delta(x-x^{\prime}),\\  \\     
-ik\mathcal{G}_{-}^{00}+m_{-}\mathcal{G}_{-}^{10}+
i\omega\mathcal{G}_{-}^{20}&=& 0,\\  \\  
\partial_x\mathcal{G}_{-}^{00} - 
i\omega\mathcal{G}_{-}^{10} + m_{-}\mathcal{G}_{-}^{20} &=& 0,
\label{g2}
\end{array}
\right.
\end{equation}

\begin{equation}
\left\{
\begin{array}{ccc}
\partial_x\mathcal{G}_{-}^{22}-
ik\mathcal{G}_{-}^{12}+m_{-}\mathcal{G}_{-}^{02}&=&0,\\  \\      
i\omega\mathcal{G}_{-}^{22}+m_{-}\mathcal{G}_{-}^{12}-
ik\mathcal{G}_{-}^{02}&=&
0,\\  \\   m_{-}\mathcal{G}_{-}^{22}-i\omega\mathcal{G}_{-}^{12}+
\partial_x\mathcal{G}_{-}^{02}&=&
\delta(x-x^{\prime}),
\label{g3}
\end{array}
\right.
\end{equation}

\begin{equation}
\left\{
\begin{array}{ccc}
ik\mathcal{G}_{+}^{01}+m_{+}\mathcal{G}_{+}^{11}-
i\omega\mathcal{G}_{+}^{21}&=&\delta(x-x^{\prime}),\\  \\      
m_{+}\mathcal{G}_{+}^{01}+ik\mathcal{G}_{+}^{11}-
\partial_x\mathcal{G}_{+}^{21}&=&
0,\\  \\  -\partial_x\mathcal{G}_{+}^{01}+
i\omega\mathcal{G}_{+}^{11}+ m_{+}\mathcal{G}_{+}^{21}&=& 0,
\label{s1}
\end{array}
\right.
\end{equation}

\begin{equation}
\left\{
\begin{array}{ccc}
-m_{+}\mathcal{G}_{+}^{00}-ik\mathcal{G}_{+}^{10}+
\partial_x\mathcal{G}_{+}^{20}&=&\delta(x-x^{\prime}),\\  \\      
ik\mathcal{G}_{+}^{00}+m_{+}\mathcal{G}_{+}^{10}-
i\omega\mathcal{G}_{+}^{20}&=& 0,\\  \\  
-\partial_x\mathcal{G}_{+}^{00}+
i\omega\mathcal{G}_{+}^{10}+m_{+}\mathcal{G}_{+}^{20}&=& 0,
\label{s2}
\end{array}
\right.
\end{equation}

\begin{equation}
\left\{
\begin{array}{ccc}
-\partial_x\mathcal{G}_{+}^{22}+
ik\mathcal{G}_{+}^{12}+m_{+}\mathcal{G}_{+}^{02}&=&0,\\  \\      
-i\omega\mathcal{G}_{+}^{22}+m_{+}\mathcal{G}_{+}^{12}+
ik\mathcal{G}_{+}^{02}&=&
0,\\  \\   m_{+}\mathcal{G}_{+}^{22}+i\omega\mathcal{G}_{+}^{12}-
\partial_x\mathcal{G}_{+}^{02}&=&
\delta(x-x^{\prime}).
\label{s3}
\end{array}
\right.
\end{equation}

The above equations are explicitly solved in the following 
for the two specific BC that we consider: for a perfect conductor (PC) and
for  a magnetically permeable (MP) boundaries, respectively.

\subsection{The Casimir force for PC boundaries}

We now describe the mapping between the original BC that can be
imposed on the original vector field $A_\mu$ of the MPCS model with
the ones imposed on the fields $f_{\mu}$ and $g_{\mu}$. The Casimir
effect follows from Eq.~(\ref{pressaoCasimirFG}).  We will first
consider PC at the boundaries, which can be
represented mathematically by  $F_1 = 0$, where 

\begin{equation}
F_{\mu} \equiv \epsilon_{\mu \nu \gamma}\partial^{\nu}A^{\gamma}\, , 
\label{FmuDoA}
\end{equation}
is the dual of $A_\mu$.  This is a BC that could not be treated for
instance in Ref.~\cite{felipeRudnei}, due to the specific form of the
mathematical transformations used in that work, based on
scalar degrees of freedom. 

In our case, the BC $F_1 = 0$ will imply (due to Eq.~(\ref{Afg})) in 
$\epsilon_{1\nu\gamma}\partial^{\nu}f^{\gamma} = 
\epsilon_{1\nu\gamma}\partial^{\nu}g^{\gamma}$, which can be written 
in terms of the dual fields $\tilde{f}_{\mu}$ and $\tilde{g}_{\mu}$, 
associated with $f_{\mu}$ and $g_{\mu}$, respectively,

\begin{eqnarray}
\label{dualDeFmu}
\tilde{f}_{\mu} \equiv \epsilon_{\mu\nu\gamma}\partial^{\nu}f^{\gamma}\, ,
\\
\label{dualDeGmu}
\tilde{g}_{\mu}  \equiv \epsilon_{\mu\nu\gamma}\partial^{\nu}g^{\gamma}\, .
\end{eqnarray}
In terms of these fields, the BC $F_1 = 0$ implies in
$\tilde{f}_{1} = \tilde{g}_{1}$. But since the PCS models are self-dual and anti self-dual, 
$\tilde{f}_{\mu}$ and $\tilde{g}_{\mu}$ are proportional to $f_{\mu}$ and $g_{\mu}$, respectively
 (this proportionality can be obtained if we 
use the Euler-Lagrange equations for $f_{\mu}$ and $g_{\mu}$).
Thus, we can write that the BC $\tilde{f}_{1} = \tilde{g}_{1}$ implies in $f_{1} = -g_{1}$ 
at the boundaries. Using then Eqs.~(\ref{gGJ}) and (\ref{fGJ}) we obtain

\begin{equation} 
\left[\int G_{-}^{1\alpha}(X,  X^{\prime})J_{(-)\alpha}(X')d X^{\prime} \right]\Bigr|_{x=0\,{\rm and}\, x=a} =  
-\left[ \int G_{+}^{1\alpha}(X, X^{\prime})J_{(+)\alpha}(X')d X^{\prime}\right]\Bigr|_{x=0\,{\rm and}\, x=a} \,.
\label{F1Null_GmaisGmenos}
\end{equation}  
Since the sources $J_{(-)\alpha}(X')$ and $J_{(+)\alpha}(X')$ are arbitrary, Eq.~(\ref{F1Null_GmaisGmenos}) implies that

\begin{equation}
G_{-}^{1\alpha}(X,  X^{\prime})\Bigr|_{x=0\,{\rm and}\, x=a} = 
G_{+}^{1\alpha}(X,  X^{\prime})\Bigr|_{x=0\,{\rm and}\, x=a} = 0\, .
\label{F1Null_GmaisGmenos-2}
\end{equation}

Note that when taking the BC, we are interested only in the limit $X \to X^{\prime}$ of 
${G_{\pm}^{\alpha\beta}(X, X^{\prime})}$,
such that we can take for instance $\exp[-i\omega(t-t')] = \exp[ik(y-y')] = 1$, e.g., 
in Eq.~(\ref{transformadaDeGmenos}). Then, Eq.~(\ref{F1Null_GmaisGmenos-2}), when expressed in 
terms of its {}Fourier transform, like in Eq.~(\ref{transformadaDeGmenos}), gives that we 
can write the BC equivalently as

\begin{equation}
\label{newBC1}
\mathcal{G}_{\pm}^{1\alpha}(k,\omega,x,x^{\prime})\Bigr|_{x=0\,{\rm and}\, x=a} = 0\, . 
\end{equation}
Hence, we note that in the present case, due to the BC, only 
$\mathcal{G}_{\pm}^{00}$ and $\mathcal{G}_{\pm}^{22}$ will contribute to the Casimir 
force $f$, Eq.~(\ref{T11_exp1}). To find the required functions, we use the 
standard method of continuity and also consider a  notation similar to the one used in
Ref.~\cite{Milton_1990} for convenience. Thus, we define:

\begin{eqnarray}
\kappa_{\pm}^{2} & = & \omega^2 - k^2 -
m_{\pm}^2,  \label{kappa_pm}  \\ \nonumber \\  ss_{\pm} & = & \sin (
\kappa_{\pm}\, x_{<}) \sin [\kappa_{\pm} (x_{>} -
  a)], \label{ss}\\  \nonumber \\   cc_{\pm} & = & \cos ( \kappa_{\pm}\, x_{<}) \cos
      [\kappa_{\pm} (x_{>} - a)], \label{cc}
\end{eqnarray}

\begin{equation}
sc_{\pm} = \left\{
\begin{array}{cc}
\sin (\kappa_{\pm}\, x) \cos [ \kappa_{\pm} (x^{\prime}- a)], &
\mathrm{if}\, x < x^{\prime}, \\ \\ \cos(\kappa_{\pm}\, x^{\prime})
\sin [\kappa_{\pm} (x - a)], & \mathrm{if}\, x > x^{\prime},
\end{array}
\right.
\label{sc}
\end{equation} 

\begin{equation}
cs_{\pm} = \left\{
\begin{array}{cc}
\cos( \kappa_{\pm}\, x) \sin [\kappa_{\pm} (x^{\prime}- a)], &
\mathrm{if}\, x < x^{\prime}, \\ \\ \sin( \kappa_{\pm}\,
x^{\prime})\ \cos[ \kappa_{\pm} (x - a)], & \mathrm{if}\, x >
x^{\prime},
\end{array}
\right.
\label{cs}
\end{equation} 
where $x_{>}$ ($x_{<}$) is the greater (smaller) value in the set
$\left\{x,x^\prime \right\}$.

To determine $\mathcal{G}_{\pm}^{22}$, it is useful to write it
in terms of $\mathcal{G}_{\pm}^{1\alpha}$, over which the BC is imposed directly. Using 
Eqs.~(\ref{g3}) and (\ref{s3}), we obtain:

\begin{equation}
\label{G22maisMenos_ini}
\mathcal{G}_{\pm}^{22}(k,\omega,x,x^{\prime}) = \frac{i}{k^2 - {\omega}^2}\left[
k\, \partial_x\mathcal{G}_{\pm}^{12}(k,\omega,x,x^{\prime}) 
- \mathcal{G}_{\pm}^{12}(k,\omega,x,x^{\prime})\,{\omega}\, m_{\pm} + 
\frac{k^2}{m_{\pm}^2}\delta(x - x^{\prime})  \right]\, .
\end{equation}

We can drop the spatial Dirac delta-function in Eq.~(\ref{G22maisMenos_ini}), since it gives 
no contribution to $\mathcal{G}_{\pm}^{22}$ (we are considering $x \neq x^{\prime}$).  
Note that dropping the spatial Dirac delta-function
corresponds physically to a renormalization, where an infinite
contribution proportional to $\delta(0)$, when evaluating the Green's
function at the same point, is removed from the Casimir force.  While
this procedure is perfectly fine for the present type of (rigid)
BC and the Casimir force is independent of this
renormalization process, the reader should be aware that this simple
renormalization procedure may not work  for other type of BC. 
{}For instance, it is known that for other types of
geometries (like circular BC, or including the case of  smooth
backgrounds), when computing the Casimir energy a special care must be
taken with this renormalization procedure,  as shown in details in
Refs.~\cite{Graham:2002xq,Graham:2002fw}. Physically,  the restriction
to the use of this BC approach to Casimir problems is related to  the
physical role of the BC: A real material at the boundaries  cannot
constrain all  modes of a field, as may be assumed in the BC approach.
In reality, the material that produces the BC should be modeled by
suitable interactions,  and the divergences must be removed by
counterterms for these interactions; the   renormalization group then
ensures that the predictive power of the theory  is not lost through
the subtraction\footnote{We thank the anonymous referee for  making
the above remark concerning the renormalization procedure and the
issues involved when computing the Casimir effect.}.

Next, we have to find a PDE for $\mathcal{G}_{\pm}^{12}$ subjected to the 
BC $\mathcal{G}_{\pm}^{12} = 0$ and to use this result in Eq.~(\ref{G22maisMenos_ini}). 
With this aim, we use again Eqs.~(\ref{g3}) and (\ref{s3}), obtaining

\begin{equation}
\label{eqParaG12maismenos}
\left(\partial_x^2 + \kappa_{+}^{2} \right) \mathcal{G}_{\pm}^{12}(k,\omega,x,x^{\prime}) =  
i \left( \frac{ k }{m_{\pm}} \partial_x \mp \omega
\right)\delta(x-x')\, .
\end{equation} 

We use the discontinuity method to solve Eq.~(\ref{eqParaG12maismenos}), 
obtaining

\begin{equation}
\label{solucaoParaG12maismenos}
\mathcal{G}_{\pm}^{12}(k,\omega,x,x^{\prime}) = -\frac{i}{\sin
  (a\kappa_{\pm})}\left( 
\frac{k}{m_{\pm}}\, sc_{\pm}\, \pm\,  \frac{\omega}{\kappa_{\pm}}\, ss_{\pm} \right)\, .
\end{equation}

By substituting Eq.~(\ref{solucaoParaG12maismenos})
in Eq.~(\ref{G22maisMenos_ini}), it follows that

\begin{equation}
\label{solucaoParaG22maisMenos}
\mathcal{G}_{\pm}^{22}(k,\omega,x,x^{\prime}) = 
\frac{k^2  \kappa_{\pm}^2\, 
  cc_{\pm}\, + \,  {\omega}^2 m_{\pm}^2\,  ss_{\pm}\, \pm\,  k\omega\kappa_{\pm} m_{\pm}\,  
  (cs_{\pm} + sc_{\pm})} {\left(k^2 - \omega^2 \right)\,  m_{\pm}\,  \kappa_{\pm}\, 
  \sin({a\, \kappa_{\pm}})}.
\end{equation}

Next, we follow an analogous procedure to find $\mathcal{G}_{\pm}^{00}$. 
{}First, we use Eqs.~(\ref{g2}) and (\ref{s2}) to write
these functions in terms of $\mathcal{G}_{\pm}^{10}$:

\begin{equation}
\label{G00emTermosDeG10-maismenos}
\mathcal{G}_{\pm}^{00}(k,\omega,x,x^{\prime}) = \frac{i}{k^2 - {\omega}^{2}}
\left[\pm\frac{k}{m_{+}}\left(\omega^{2} - k^2\right)
+ \omega\partial_x \pm \frac{k}{m_{+}} \partial_x^2
\right]\mathcal{G}_{+}^{10}(k,\omega,x,x^{\prime})\, .
\end{equation}

Using Eqs.~(\ref{g2}) and (\ref{s2}), we obtain

\begin{equation}
\label{EqParaG10maismenos}
\left(\kappa_{\pm}^2 + \partial_{x}^2
\right)\mathcal{G}_{\pm}^{10}(k,\omega,x,x^{\prime}) = 
i\left(\frac{\omega}{m_{\pm}}\partial_{x} \mp k\right)\delta(x - x^{\prime})\, .
\end{equation}

{}From Eq.~(\ref{EqParaG10maismenos}), we find

\begin{equation}
\label{G10maismenos}
\mathcal{G}_{\pm}^{10}(k,\omega,x,x^{\prime}) = 
\frac{\mp i}{\sin(\kappa_{\pm}\, a)}\left(\frac{k}{\kappa_{\pm}}\, ss_{\pm} 
\mp \frac{\omega}{m_{\pm}}\, sc_{\pm} \right)\, .
\end{equation}

Substituting Eq.~(\ref{G10maismenos}) in  
Eq.~(\ref{G00emTermosDeG10-maismenos}), we find

\begin{equation}
\label{solucaoParaG00maismenos}
\mathcal{G}_{\pm}^{00}(k,\omega,x,x^{\prime})  =  \frac{ {\omega}^2 \kappa_{\pm}^2\, 
  cc_{\pm}\,  +\,  k^2  m_{\pm}^2\,  ss_{\pm}\, \pm\,  m_{\pm}\kappa_{\pm}\, k\, \omega\, 
  (cs_{\pm} +  sc_{\pm})} {\left(k^2 - \omega^2\right)\,  m_{\pm}\, \kappa_{\pm}\, 
  \sin({a\, \kappa_{\pm}})}\, .
\end{equation}

Inserting the expressions for $\mathcal{G}_{\pm}^{00}$ and
$\mathcal{G}_{\pm}^{22}$, together with $\mathcal{G}_{\pm}^{11} = 0$,  in
Eq.~(\ref{T11_exp1}), we can write the Casimir force for the PC BC
case as

\begin{equation}
\label{fc0}
f_{PC} =  \left( \left\langle\,  T^{11}_{-}\, \right\rangle +
\left\langle\,  T^{11}_{+}\, \right\rangle \right)\Bigr|_{x=0\,{\rm and}\, x=a} =
\frac{i}{2}\int\frac{d\omega}{2\pi} \int\frac{dk}{2\pi}
\left[\kappa_{+}\cot(a\, \kappa_{+}) + \kappa_{-}\cot(a\, \kappa_{-})\right]\, .
\end{equation}

The integrals appearing in Eq.~(\ref{fc0}) can be evaluated 
in an analogous fashion as in Ref.~\cite{Milton_1990}.
{}First, we make a complex rotation $\omega \to i\zeta$, where $\zeta$
is real (this is possible since there are no poles in the first and in
the third quadrants). The effect of this rotation is to turn
$\kappa_{\pm} \equiv ( \omega^2 - k^2 - m_{\pm}^2)^{1/2}$ into a
purely complex variable. Then we can redefine it as $\kappa_{\pm}
= i\lambda_{\pm}$, where $\lambda_{\pm} = \sqrt{\zeta^2 + k^2 +
m_\pm^2} $ is a real variable. Then,
using the relation

\begin{equation}
\cot(\kappa_{\pm} a) = -i\left[ 1 + \frac{2}{\exp(2\,\lambda_{\pm}\,a)
    - 1} \right],
\end{equation}
we can rewrite Eq.~(\ref{fc0}) as an integral defined entirely in
the real $(\zeta, k)$ plane, where

\begin{equation}
\left\langle\,  T^{11}_{\pm}\, \right\rangle\Bigr|_{x=0\,{\rm and}\, x=a} =
- \int\frac{d\zeta}{2\pi}\, \int\frac{dk}{2\pi}\frac{\lambda_{\pm}}{\left[
    \exp(2\,\lambda_{\pm}\,a) - 1\right]}\,.
\label{T11pm_b}
\end{equation}

We can also write Eq.~(\ref{T11pm_b}) in terms of polar coordinates
$(r,\, \phi)$, defined by

\begin{eqnarray}
\zeta & = & r\cos\phi\, ,
\label{zetarphi}\\
k & = & r\sin\phi\, .
\label{krphi}
\end{eqnarray}

Substituting Eqs.~(\ref{zetarphi}) and (\ref{krphi}) in Eq.~(\ref{T11pm_b})
and performing the integration over $\phi$, we obtain,

\begin{equation}
\left\langle\,  T^{11}_{\pm}\, \right\rangle \Bigr|_{x=0\,{\rm and}\, x=a} 
=
-\int_0 ^\infty \frac{\ dr}{2\pi}  
\frac {r \sqrt {m_{\pm}^{2}+ r^2} }
{ \left[\exp\left(2a\,\sqrt{m_{\pm}^{2}+{r}^{2}} \right) - 1 \right]} 
= -\int_{m_{\pm}}^{\infty} \frac{\ d\lambda}{2\pi} \,  
\frac{ \lambda^2}{\left[ \exp(2\,\lambda\,a) - 1 \right] }\, ,
\label{T11pm_d}
\end{equation}
where to obtain the last expression on the right-hand side in
Eq.~(\ref{T11pm_d}), we have made a change of integration variables,
using $\lambda^2 = r^2 + m_{\pm}^2$. {}From this equation, we can write the Casimir 
force for the case of PC boundaries as (when making the
change of variables: $z = 2\,\lambda\,a$):

\begin{equation}
\label{ffinal}
f_{PC} =-\frac{1}{16\pi a^3}\left[\int_{2am_{-}}^{\infty}dz \frac{z^2}{e^z-1}
+\int_{2am_{+}}^{\infty}dz \frac{z^2}{e^z-1}\right].
\end{equation}

\subsection{The Casimir force for perfect MP boundaries}

{}Following an analogous derivation as outlined in the previous
subsection, we now derive the Casimir force for the case of perfect
MP lines.  The same mapping
relating the MPCS with a doublet made of a self dual and an anti-self dual 
PCS models is, of course, still applicable, as also the system of PDE, 
Eqs.~(\ref{g1})-(\ref{s3}), derived previously.
Perfect MP lines at the boundaries are represented
by the BC $F_0 = 0$. This BC, in turn, can be represented in terms of
$\mathcal{G}_{\pm}^{\mu\nu}$, analogously to what we have done in the previous 
subsection to obtain the BC given in Eq.~(\ref{newBC1}). Thus, we find that we
can write the present BC as:

\begin{equation}
\label{newBC2}
\mathcal{G}_{\pm}^{0\alpha}\Bigr|_{x=0\,{\rm and}\, x=a} = 0\, . 
\end{equation}

Using Eqs.~(\ref{T11_exp1})-(\ref{newBC2}), we can see that $\mathcal{G}_{\pm}^{00}$ do not
contribute to the Casimir force at the boundaries. Hence, we only need to obtain
$\mathcal{G}_{\pm}^{11}$ and $\mathcal{G}_{\pm}^{22}$.

Using an analogous procedure as the one used in the previous subsection, and noting 
that the BC is imposed on $\mathcal{G}_{\pm}^{0\alpha}$, we
first find a relation between $\mathcal{G}_{\pm}^{22}$ and
$\mathcal{G}_{\pm}^{02}$. Analogously, we need to find a relation between 
$\mathcal{G}_{\pm}^{11}$and 
$\mathcal{G}_{\pm}^{01}$.  {}For example, for 
$\mathcal{G}_{+}^{22}$, we can write (and again dropping a space Dirac delta-function
for the same reason explained in the previous subsection):

\begin{equation}
\label{relacaoG22-G02}
\mathcal{G}_{\pm}^{22}(k,\omega,x,x^{\prime}) = \frac{\left(k\omega\, \mp\,  
m\partial_{x}\right)}{{\omega}^{2}
- m_{\pm}^{2} }\mathcal{G}_{\pm}^{02}(k,\omega,x,x^{\prime})\, ,
\end{equation}
and

\begin{equation}
\label{eqParaG02fg}
\left(\partial_x^2 + \kappa_{\pm}^{2} \right) \mathcal{G}_{\pm}^{02}(k,\omega,x,x^{\prime}) 
=  -\left(\frac{k\omega }{m_{\pm}} \pm \partial_x  
\right)\delta(x-x')\, ,
\end{equation} 
which has the solution

\begin{equation}
\label{solucaoParaG02fg}
\mathcal{G}_{\pm}^{02}(k,\omega,x,x^{\prime}) = -\frac{(\omega k\, ss_{\pm}\, \mp\, 
m_{\pm}\, \kappa_{\pm}\, sc_{\pm})}
{m_{\pm}\, \kappa_{\pm}\, \sin(a\, \kappa_{\pm})} \, . 
\end{equation}
Hence,

\begin{equation}
\label{solucaoParaG22fg}
\mathcal{G}_{\pm}^{22}(k,\omega,x,x^{\prime}) =  
\frac{m_{\pm}^2 \kappa_{\pm}^2\, cc_{\pm} 
\, +\, k^2\,  {\omega}^2\, ss_{\pm}\,  \mp\,  m_{\pm}\, \kappa_{\pm}\, k\, 
\omega(cs_{\pm} +  sc_{\pm})}{\left(m_{\pm}^2 - 
  {\omega}^2 \right)\, m_{\pm}\,  \kappa_{\pm}\,  \sin(a\, \kappa_{\pm})}\, . 
\end{equation}

The procedure to find $\mathcal{G}_{\pm}^{11}$ is completely analogous,
leading to the result

\begin{equation}
\label{solucaoParaG11fg}
\mathcal{G}_{\pm}^{11}(k,\omega,x,x^{\prime}) =  
\frac{\omega_{\pm}^2 \kappa_{\pm}^2\, cc_{\pm} 
\, +\, k^2\, m_{\pm}^2\, ss_{\pm}\,  \mp\,  m_{\pm}\, \kappa_{\pm}
\, k\omega\, (cs_{\pm} +  sc_{\pm})}{\left(m_{\pm}^2 - 
  {\omega}^2 \right)\, m_{\pm}\, \kappa_{\pm}\, \sin(a\kappa_{\pm})}\,.
\end{equation}

Using the above expressions for $\mathcal{G}_{\pm}^{11}$ and
$\mathcal{G}_{\pm}^{22}$, together with $\mathcal{G}_{\pm}^{00} = 0$, 
in Eq.~(\ref{T11_exp1}), it can be easily verified that this results
again in the same Casimir force as derived in the previous subsection, Eq.~(\ref{fc0}), 
leading also to Eq.~(\ref{ffinal}), i.e.,
$f_{MP} = f_{PC}$. In the next two sections we try to understand this rather
surprising result.

\section{Casimir force from the mapping between
the MPCS model and a doublet of MCS models}
\label{sec4}

In the previous section we have obtained that the Casimir force for the MPCS model
is independent of the two type of BC considered, i.e., for PC and MP lines at the boundaries.
In this section we verify whether this result is not a consequence of the particular
mapping that we have used, involving the relation of the MPCS model with a self dual and an anti-self
dual PCS models, described in section \ref{2PCS}. {}For this, we use the second
relationship discussed in section \ref{2MCS}, relating the MPCS model with a doublet
of MCS models, expressed by Eqs.~(\ref{LmenosP}) and (\ref{LmaisQ}). 

\subsection{Casimir force for perfect MP boundaries}

We here specialize to 
the case of the perfect MP BC $F_0 = 0$. This analysis is made easier by the fact that
the Casimir force for a MCS model under the BC $F_0 = 0$ was already studied in 
Ref.~\cite{DanEdneyFelSil}. The results found in that
reference can be easily extended to the Lagrangian densities given by
Eqs.~(\ref{LmenosP}) and (\ref{LmaisQ}), as we show below.

The Casimir force for the MPCS model can be obtained from the sum of the 11 
component of the total energy-momentum tensor determined from the Lagrangian
densities (\ref{LmenosP}) and (\ref{LmaisQ}), i.e.,

\begin{equation}
\label{fc_test1}
f = \left[\left\langle\, T^{11}_{(P)}\, \right\rangle  +
\left\langle\,  T^{11}_{(Q)}\, \right\rangle \right]\Bigr|_{x=0\,{\rm and}\, x=a}\, , 
\end{equation}
where $T^{11}_{(P)}$ and $T^{11}_{(Q)}$ are the 11 component of the total 
energy-momentum tensor associated with $\tilde{\mathcal{L}}_{-}(P)$ and
$\tilde{\mathcal{L}}_{+}(Q)$, Eqs.~(\ref{LmenosP}) and (\ref{LmaisQ}), respectively. 

Let us first consider $T^{11}_{(P)}$. Our considerations can be easily extended 
to $T^{11}_{(Q)}$. Using analogous procedures as the ones used in the 
previous section, we can write

\begin{equation}
\left\langle\, T^{\mu\nu}_{(P)} \, \right\rangle \Bigr|_{x=0\,{\rm and}\, x=a}= 
\left( \left\langle\, \tilde{P}^{\mu}\tilde{P}^{\nu} \, \right\rangle
- \frac{1}{2}\eta^{\mu\nu}\left\langle\, \tilde{P}_{\alpha}\tilde{P}^{\alpha}
\, \right\rangle  \right)\Bigr|_{x=0\,{\rm and}\, x=a} \, ,
\label{TmunuP}
\end{equation}
where $\tilde{P}^{\mu} = \epsilon^{\mu\alpha\beta}\partial_{\alpha}P_{\beta}$.
Analogously, we define $\tilde{Q}^{\mu} = \epsilon^{\mu\alpha\beta}\partial_{\alpha}Q_{\beta}$.
The VEV $\left\langle\, \tilde{P}^{\mu}\tilde{P}^{\nu}  \, \right\rangle$ can be obtained from 
$\left\langle\, \tilde{P}^{\mu}(X)\tilde{P}^{\nu}(X^{\prime}) 
\, \right\rangle$ as

\begin{equation}
\label{PtilPtil}
\left\langle\, \tilde{P}^{\mu}\tilde{P}^{\nu}
\, \right\rangle = \lim_{X \to X^{\prime}}
\left\langle\, \tilde{P}^{\mu}(X)\tilde{P}^{\nu}(X^{\prime}) 
\, \right\rangle \, ,
\end{equation}
and $\left\langle\, \tilde{P}^{\mu}(X)\tilde{P}^{\nu}(X^{\prime}) 
\, \right\rangle$ can be related to the Green's function 
$G_{(P)}^{\mu\rho}$ for $\tilde{P}^{\mu}$, as we show below. 

We know that $G_{(P)}^{\mu\rho}$ can be obtained from the Euler-Lagrange equation associated with
$\tilde{\mathcal{L}}_{-}(P)$, written in terms of $\tilde{P}^{\mu}$, plus a 
source term:

\begin{equation}
\tilde{\mathcal{L}}_{-}(P) = -\frac{1}{2}\tilde{P}_{\mu}\tilde{P}^{\mu}
+\frac{1}{2} m_{-} \tilde{P}^{\mu}P_{\mu} + J^{\mu}P_{\mu}\, .
\label{LmenosPtil}
\end{equation}     

Considering the equation of motion 

\begin{equation}
- \epsilon^{\mu\alpha\beta}\partial_{\alpha}\tilde{P}_{\beta} 
+ m_{-}\tilde{P}^{\mu} + J^{\mu} = 0,
\label{eqMotionPtil}
\end{equation}     
with formal solution

\begin{equation}
\tilde{P}^{\mu} = \int G_{(P)}^{\mu\rho}(X,X^\prime)J_{\rho}
(X^{\prime}) dX^{\prime}\, ,
\label{PtilGP}
\end{equation}     
we obtain the differential equation satisfied by 
$G_{(P)}^{\mu\rho}(X,X^\prime)$:

\begin{equation}
\label{eqForGreenP}
\left( \epsilon_{\nu\alpha\beta}\partial^{\alpha} 
- m_{-}\eta_{\nu\beta}\right)G_{(P)}^{\beta\rho}(X,X^\prime) = 
\delta^{\rho}_{\nu}\delta(X - X^{\prime})\, .
\end{equation}
We then solve Eq.~(\ref{eqForGreenP}) to find
the functions $G_{(P)}^{\beta\rho}(X,X^\prime)$ that will be 
necessary to compute $\left\langle\, T^{\mu\nu}_{(P)} \, \right\rangle$
in Eq.~(\ref{TmunuP}). {}First, we need a relation between
$G_{(P)}^{\beta\rho}(X,X^\prime)$ and 
$\left\langle\, \tilde{P}^{\beta}(X)\tilde{P}^{\rho}(X^{\prime}) 
\, \right\rangle$. {}For this purpose, we 
consider the propagator for $P^{\mu}$,

\begin{equation}
\label{propagatorP}
\Delta^{\beta\rho}(X,X^\prime) = i\left\langle\, P^{\beta}(X)P^{\rho}(X^{\prime}) 
\, \right\rangle\, ,
\end{equation}
where $\left\langle\, P^{\beta}(X)P^{\rho}(X^{\prime}) 
\, \right\rangle$ is the Green's function 
for $P^{\mu}$, which can be obtained directly from 
the equation of motion generated by Eq.~(\ref{LmenosP}),
when including a source term $J^{\mu}P_{\mu}$, as above.
Hence, we can write~\cite{Milton_1990}:

\begin{equation}
\label{GDeltaPtilP}
G_{(P)}^{\beta\rho}(X,X^\prime) = 
\epsilon^{\beta\alpha\nu}\partial_{\alpha}{\Delta_{\nu}}^{\rho}(X,X^\prime)
= i\left\langle\, \tilde{P}^{\beta}(X)P^{\rho}(X^{\prime}) 
\, \right\rangle\, .
\end{equation}

{}From Eq.~(\ref{GDeltaPtilP}), we obtain

\begin{equation}
\label{PtilPtilG}
\left\langle\, \tilde{P}^{\beta}(X)\tilde{P}^{\rho}(X^{\prime}) 
\, \right\rangle = -i\epsilon^{\rho\alpha\gamma}\, \partial^{\prime}_{\alpha}
{{G_{(P)}^{\beta}}}_{\gamma}(X,X^\prime)\, .
\end{equation}

Using Eq.~(\ref{TmunuP}), we can write

\begin{equation}
\left\langle\, T^{11}_{(P)} \, \right\rangle \Bigr|_{x=0\,{\rm and}\, x=a}= 
\frac{1}{2}\left( \left\langle\, P^{0}P^{0} \, \right\rangle
+ \left\langle\, P^{1}P^{1} \, \right\rangle
- \left\langle\, P^{2}P^{2} \, \right\rangle
  \right)\Bigr|_{x=0\,{\rm and}\, x=a}\, ,
\label{TmunuPbar}
\end{equation}
where

\begin{equation}
\label{propagatorP_bar}
\left\langle\, P^{\mu}P^{\nu} 
\, \right\rangle\Bigr|_{x=0\,{\rm and}\, x=a} = -i\lim_{X \to X^{\prime}}
{\epsilon^{\nu}}_{\lambda\rho}\partial^{\prime\lambda}
{G_{(P)}^{\mu\rho}}(X,X^\prime)\Bigr|_{x=0\,{\rm and}\, x=a}\, ,
\end{equation}
and ${G_{(P)}^{\mu\rho}}(X,X^\prime)$ satisfies

\begin{equation}
\label{PDEforGbar}
\left( \epsilon_{\nu\alpha\beta}\partial^{\alpha} 
- m_{-}\eta_{\nu\beta}\right){G_{(P)}^{\beta\rho}}(X,X^\prime) = 
\delta^{\rho}_{\nu} \delta(X - X^{\prime})\, .
\end{equation}

Considering the {}Fourier transform of ${G_{(P)}^{\mu\rho}}$
(with respect to $t$ and $y$), 

\begin{equation}
\label{Gcal_bar}
{G_{(P)}^{\mu\rho}}(X,X^\prime) = 
\int\frac{d\omega}{2\pi} e^{-i\omega(t-t')}
\int\frac{dk}{2\pi} e^{ik(y-y')} 
{{\cal G}_{(P)}^{\mu\rho}}(k,\omega,x,x^{\prime})\,,
\end{equation}
we can write, using Eqs.~(\ref{TmunuPbar}) and (\ref{propagatorP_bar}), that

\begin{equation}
\left\langle\, T^{11}_{(P)} \, \right\rangle \Bigr|_{x=0\,{\rm and}\, x=a} = 
\lim_{X \to X^{\prime}}\int\frac{d\omega}{2\pi}\, e^{-i\omega(t-t')} 
\int\frac{dk}{2\pi}\, e^{ik(y-y')}t^{11}_{(P)}\Bigr|_{x=0\,{\rm and}\, x=a}\, ,
\label{T11Pbar2}
\end{equation}
where

\begin{equation}
t^{11}_{(P)} = 
\frac{i}{2}\frac{\partial}{\partial x^{\prime}}
\left({{\cal G}_{(P)}^{02}} - {{\cal G}_{(P)}^{20}} \right) -
\frac{k}{2}
\left({{\cal G}_{(P)}^{01}} + {{\cal G}_{(P)}^{10}}\right)
+ \frac{\omega}{2}
\left({{\cal G}_{(P)}^{12}} + {{\cal G}_{(P)}^{21}}\right)\, .
\label{t11P}
\end{equation}

The required functions ${{\cal G}_{(P)}^{\mu\nu}}$ can be obtained from
Eqs.~(\ref{PDEforGbar}) and (\ref{Gcal_bar}), analogously to what we have done
in the previous sections. We can write

\begin{equation}
\left\{
\begin{array}{ccc}
- i k {{\cal G}_{(P)}^{01}} + m_{-} {{\cal G}_{(P)}^{11}} + 
i \omega {{\cal G}_{(P)}^{21}} &=&
\delta(x-x^{\prime})\, , \\ \\ 
m_{-}{{\cal G}_{(P)}^{01}} -i k {{\cal G}_{(P)}^{11}} + 
\partial_{x}{{\cal G}_{(P)}^{21}} &=&0\, ,
\\ \\ 
\partial_{x}{{\cal G}_{(P)}^{01}} - 
i \omega {{\cal G}_{(P)}^{11}} + m_{-}{{\cal G}_{(P)}^{21}} &=& 0\, ,
\label{sp1}
\end{array}
\right.
\end{equation}

\begin{equation}
\left\{
\begin{array}{ccc}
- m_{-} {{\cal G}_{(P)}^{00}} + i k {{\cal G}_{(P)}^{10}} - 
\partial_{x}{{\cal G}_{(P)}^{20}} &=& \delta (x-x^{\prime})\, ,
\\ \\ 
- i k{{\cal G}_{(P)}^{00}} + m_{-} {{\cal G}_{(P)}^{10}} + 
i \omega {{\cal G}_{(P)}^{20}} &=& 0\, ,
\\ \\ 
\partial_{x}{{\cal G}_{(P)}^{00}} - 
i \omega {{\cal G}_{(P)}^{10}} + m_{-} {{\cal G}_{(P)}^{20}} &=& 0\, ,
\label{sp2}
\end{array}
\right.
\end{equation}

\begin{equation}
\left\{
\begin{array}{ccc}
\partial_{x} {{\cal G}_{(P)}^{22}} - 
i k {{\cal G}_{(P)}^{12}} + m_{-}{\cal G}_{(P)}^{02} &=& 0\, ,
\\ \\ 
i \omega{\cal G}_{(P)}^{22} + m_{-} {\cal G}_{(P)}^{12} - 
i k {\cal G}_{(P)}^{02} &=& 0\, ,
\\ \\ 
m_{-} {\cal G}_{(P)}^{22}- i \omega{\cal G}_{(P)}^{12} + 
\partial_{x}{\cal G}_{(P)}^{02} &=& \delta(x-x^{\prime})\,.
\label{sp3}
\end{array}
\right.
\end{equation}

We note that the above equations are the same ones as those treated in 
Ref.~\cite{DanEdneyFelSil} and, also, the forms of $T^{11}_{(P)}$ and $t^{11}_{(P)}$ are 
analogous to the ones derived in that reference. 
In the present case, where we are considering the BC $F_0 = 0$, using Eq.~(\ref{APQ}), we obtain that
$\sqrt{m_{-}}\tilde{P}_{0}(X) = \sqrt{m_{+}}\tilde{Q}_{0}(X)$ at the boundaries. Hence, 
using an analogous procedure as used to obtain Eq.~(\ref{newBC1}) and considering Eq.~(\ref{PtilGP}),
we can write the BC in the present case as

\begin{equation}
\label{F0Null_GPGQ_bar}
{\cal G}_{(P)}^{0\rho}(k,\omega,x,x^{\prime}) \Bigr|_{x=0\,{\rm and}\, x=a} =
{{\cal G}_{(Q)}^{0\rho}}(k,\omega,x,x^{\prime}) \Bigr|_{x=0\,{\rm and}\, x=a} = 0\, .
\end{equation}

Hence, we conclude from Eqs.~(\ref{T11Pbar2}), (\ref{t11P}) and (\ref{F0Null_GPGQ_bar}) that
we only need to find $\frac{\partial}{\partial x^{\prime}}{{\cal G}_{(P)}^{20}}$, 
$\frac{\partial}{\partial x^{\prime}}{{\cal G}_{(P)}^{02}}$, ${{\cal G}_{(P)}^{10}}$,
${{\cal G}_{(P)}^{12}}$ and ${{\cal G}_{(P)}^{21}}$  to compute
$\left\langle\, T^{11}_{(P)} \, \right\rangle$ at $x=0$ and $x=a$.
As already commented in the Introduction, we note that the number of functions 
that we need to find, in the case of the mapping treated in this section,
is greater than the number of required functions in the case considered in the previous section
(where we considered the mapping between the MPCS model and the two PCS models).

The solutions to Eq.~(\ref{PDEforGbar}), considering the BC
given in Eq.~(\ref{F0Null_GPGQ_bar}), are given by~\cite{DanEdneyFelSil}:

\begin{eqnarray}
{{\cal G}_{(P)}^{21}}(k,\omega,x,x^{\prime}) & = &  
\frac{- i \omega}{(\omega^2 - m_{-}^2)\sin (a\kappa_{-})} \left(\frac{k^2}{\kappa_{-}} 
ss_{-} + \frac{k \omega}{m_{-}} sc_{-} + \frac{k m_{-}}{\omega}cs_{-} + 
\kappa_{-} cc_{-} \right)\, , \label{Gcal_bar21}\\ \nonumber \\
{{\cal G}_{(P)}^{12}}(k,\omega,x,x^{\prime}) & = &  
\frac{i \omega}{(\omega^2 - m_{-}^2)\sin (a\kappa_{-})} \left(\frac{k^2}{\kappa_{-}} 
ss_{-} + \frac{k \omega}{m_{-}} cs_{-} + \frac{k m_{-}}{\omega}sc_{-} + 
\kappa_{-} cc_{-} \right)\, , \label{Gcal_bar12}\\ \nonumber \\
{{\cal G}_{(P)}^{20}}(k,\omega,x,x^{\prime}) & = &  
\frac{-1}{\sin(a\kappa_{-})}\left(\frac{\omega k}{m_{-} \kappa_{-}}\, ss_{-}
+ cs_{-} \right) \, , \label{Gcal_bar20}\\ \nonumber \\
{{\cal G}_{(P)}^{10}}(k,\omega,x,x^{\prime}) & = &  
\frac{i}{\sin(a\kappa_{-})} \left(\frac{\omega}{m_{-}}\, cs_{-} +
\frac{k}{\kappa_{-}}\,  ss_{-}\right) \, . \label{Gcal_bar10}
\end{eqnarray}

Substituting Eqs.~(\ref{Gcal_bar21})-(\ref{Gcal_bar10}) in Eqs.~(\ref{T11Pbar2}) and (\ref{t11P}), 
we obtain

\begin{equation}
\label{T11P-final}
\left\langle\, T^{11}_{(P)} \, \right\rangle \Bigr|_{x=0\,{\rm and}\, x=a}
= \frac{i}{2}\int\frac{d\omega}{2\pi} \int\frac{dk}{2\pi}
\kappa_{-}\cot(a\, \kappa_{-})\, .
\end{equation}

The derivation of $\left\langle\, T^{11}_{(Q)} \, \right\rangle$ is completely analogous and the result 
found is

\begin{equation}
\left\langle\, T^{11}_{(Q)} \, \right\rangle \Bigr|_{x=0\,{\rm and}\, x=a}
= \frac{i}{2}\int_{-\infty}^{\infty}\frac{d\omega}{2\pi} 
\int_{-\infty}^{\infty}\frac{dk}{2\pi} \kappa_{+}\cot(a\, \kappa_{+})\, .
\label{T11Q-final}
\end{equation}

Thus, from Eqs.~(\ref{fc_test1}) and (\ref{T11P-final}), we obtain 
again Eq.~(\ref{fc0}). This confirms our previous result and at the same time
it shows that the result obtained for the Casimir force is independent
of the mapping used for the case of a MP BC.

\subsection{Casimir force for PC boundaries}

We can also use the mapping between the MPCS model and $\tilde{\mathcal{L}}_{-}(P) + 
\tilde{\mathcal{L}}_{+}(Q)$ to also confirm
our result for the Casimir force in the case of a PC BC,
$F_1 = 0$. The MCS model under this BC was considered in Ref.~\cite{Milton_1990}
and the results found there can be easily extended to the case treated here,
in the same way as we did in previous subsection.

In this case Eqs.~(\ref{T11Pbar2}) and (\ref{t11P})
still remain valid and also the PDE satisfied by ${\cal G}_{(P)}^{\mu\nu}$
and ${\cal G}_{(Q)}^{\mu\nu}$. We then have that

\begin{equation}
\left\langle\, T^{11}_{(P)} \, \right\rangle \Bigr|_{x=0\,{\rm and}\, x=a} = 
\lim_{X \to X^{\prime}}\int\frac{d\omega}{2\pi}e^{-i\omega(t-t')} 
\int\frac{dk}{2\pi}e^{ik(y-y')} t^{11}_{(P)}\Bigr|_{x=0\,{\rm and}\, x=a}\, ,
\label{T11Qbar3}
\end{equation}
where

\begin{equation}
t^{11}_{(P)} = 
\frac{1}{2i}\frac{\partial}{\partial x^{\prime}}
\left({{\cal G}_{(P)}^{02}} - {{\cal G}_{(P)}^{20}} \right) +
\frac{k}{2}
\left({{\cal G}_{(P)}^{01}} + {{\cal G}_{(P)}^{10}}\right)
+ \frac{\omega}{2}
\left({{\cal G}_{(P)}^{12}} + {{\cal G}_{(P)}^{21}}\right)\,.
\label{t11PF1Null}
\end{equation}

As in the previous cases, we can conclude that the BC $F_1 = 0$ implies in 

\begin{equation}
\label{F1Null_GPGQ_bar}
{\cal G}_{(P)}^{1\nu}(k,\omega,x,x^{\prime}) \Bigr|_{x=0\,{\rm and}\, x=a} =
{\cal G}_{(Q)}^{1\nu}(k,\omega,x,x^{\prime}) \Bigr|_{x=0\,{\rm and}\, x=a} = 0\, .
\end{equation}

To obtain the Casimir force $f$ at the boundaries, we need
$\frac{\partial}{\partial x^{\prime}}{{\cal G}_{(P)}^{02}}$, $\frac{\partial}{\partial x^{\prime}}{{\cal G}_{(P)}^{20}}$, 
${{\cal G}_{(P)}^{01}}$, ${{\cal G}_{(P)}^{21}}$ and the 
corresponding ${\cal G}_{(Q)}^{\mu\nu}$. The required
functions are now found to be given by

\begin{eqnarray}
{{\cal G}_{(P)}^{02}}(k,\omega,x,x^{\prime}) & = &  \frac{1}{\sin (a\kappa_{-})} 
\left(\frac{k m_{-} \omega}{\kappa_{-}\rho_{-}^2}\, ss_{-} + \frac{k^2}{\rho_{-}^2}\,  sc_{-} + 
\frac{\omega^2}{\rho_{-}^2}\,  cs_{-} + \frac{\omega k \kappa_{-}}{m_{-}\rho_{-}^2}\,  cc_{-} \right)\, , 
\label{Gcal_bar02_F1Null}\\ \nonumber \\
{{\cal G}_{(P)}^{20}}(k,\omega,x,x^{\prime}) & = & \frac{1}{\sin (a\kappa_{-})} 
\left(\frac{k m_{-} \omega}{\kappa_{-}\rho_{-}^2}\, ss_{-} + \frac{k^2}{\rho_{-}^2}\, cs_{-} + 
\frac{\omega^2}{\rho_{-}^2}\, sc_{-} + \frac{\omega k \kappa_{-}}{m_{-}\rho_{-}^2}\, cc_{-} \right) \, , 
\label{Gcal_bar20_F1Null}\\ \nonumber \\
{{\cal G}_{(P)}^{01}}(k,\omega,x,x^{\prime}) & = & \frac{i}{\sin(a\kappa_{-})} 
\left(\frac{k}{\kappa_{-}}\, ss_{-} +
\frac{\omega}{m_{-}}\, cs_{-}\right) \, , \label{Gcal_bar01_F1Null}\\ \nonumber \\
{{\cal G}_{(P)}^{21}}(k,\omega,x,x^{\prime}) & = &  \frac{i}{\sin(a\kappa_{-})} 
\left(\frac{\omega}{\kappa_{-}}\, ss_{-} +
\frac{k}{m_{-}}\, cs_{-}\right) \, . \label{Gcal_bar21_F1Null}
\end{eqnarray}

Using Eq.~(\ref{t11PF1Null}) and 
Eqs.~(\ref{Gcal_bar02_F1Null})-(\ref{Gcal_bar21_F1Null}), we obtain

\begin{equation}
\label{T11P-final_F1Null}
\left\langle\, T^{11}_{(P)} \, \right\rangle \Bigr|_{x=0\,{\rm and}\, x=a}
= \frac{i}{2}\int\frac{d\omega}{2\pi} \int\frac{dk}{2\pi}
\kappa_{-}\cot(a\, \kappa_{-})\,.
\end{equation}
The procedure to find
$\left\langle\, T^{11}_{(Q)} \, \right\rangle$ is again completely analogous and we do not need
to repeat it again here. The final result
that we find is once again the same one given in Eq.~(\ref{T11Q-final}). Thus, we are again lead
to the very same previous result for the Casimir force, given by
Eq.~(\ref{ffinal}).

\section{Interpreting the independence of the results for different
boundary conditions}
\label{sec5}

Casimir forces are, in general, sensible to the BC changes. However, in the previous calculations, 
we have shown that, for the MPCS model, it does not depend whether we have MP or PC BC.
In this section, we are willing to find an argument that sustains this coincidence, 
as well as to find out some others equivalent BC.
The fact that the Casimir force obtained with both the PC and MP boundaries are
the same can be understood as a consequence of the fact that the components $f_{\mu}$ 
(or $g_{\mu}$) are not independent from each other (since there are three components
$A_{\mu}$ and just two degrees of freedom). To see this interdependence more clearly, we
can use the relations obtained for the canonical momenta in the model, 

\begin{equation}
\label{momenta}
\pi_{\nu} = \frac{\partial \mathcal{L}}{\partial \dot{A}^{\nu}}\, ,
\end{equation}
where $\mathcal{L}$ is given in Eq.~(\ref{LMPCS}). The MPCS model has two constraints: 
\begin{equation}
\label{constraint1}
\pi_0 \approx 0\, 
\end{equation}
and
\begin{equation}
\label{constraint2}
\partial_i\pi_i - \frac{\mu}{4}\epsilon_{ij}\partial_i A_j - m^2 A_0 \approx 0\, ,
\end{equation}
where the "$\approx$" symbol is used to emphasize that both constraints are secondary
 and $\pi_i = F_{0i} + (\mu/4)\epsilon_{ji}A_{j}$. The second constraint,
Eq.~(\ref{constraint2}), shows us that $A_0$ is not an independent variable    
(the same can be said about $f_0$ and also for $g_0$). Indeed, using 
Eqs.~(\ref{constraint1}) and (\ref{constraint2}), we can write the 
generating functional $Z$ only in terms of $\left\{ A_{i}, \pi_{i}\right\}$
(and analogously for $f_{\mu}$ and $g_{\mu}$). 
  
Another important conclusion about the Casimir force, 
concerning the interdependence of $f_{\mu}$ and $g_{\mu}$, 
in the case of the BC $F_0 =0$, can be obtained as follows. 
Using the equations of motion for $g_{\mu}$ and $f_{\mu}$, given 
by Eqs.~(\ref{eqMotion_g}) 
and (\ref{eqMotion_f}), respectively, we can obtain~\cite{{BanerjeeKumar2}}:

\begin{eqnarray}
\label{eq2_f}
m_{+}\epsilon^{\mu\nu\gamma}\partial_{\nu}f_{\gamma} & = & -\partial_{\alpha} f^{\mu\alpha}\, ,
\\
\label{eq2_g}
m_{-}\epsilon^{\mu\nu\gamma}\partial_{\nu}g_{\gamma} & = & \partial_{\alpha} g^{\mu\alpha}\, .
\label{eqforg}
\end{eqnarray}
Thus, we find the following relations satisfied by the vector field $f^\mu$:

\begin{eqnarray}
\label{eq_f0}
m_{+}f^{0} & = & f^{21}\, ,
\\
\label{eq_f1}
m_{+}f^{1} & = & f^{20}\, ,
\\
\label{eq_f2}
m_{+}f^{2} & = & f^{01}\, ,
\\
\label{eq10_contas}
m_{+}^2f^{1} & = & \partial_{\mu}f^{1\mu}\, ,
\\
\label{Bianchi}
\partial_{\mu}f^{\mu} & = & 0\, ,
\end{eqnarray}
where $f_{\alpha\beta} = \partial_{\alpha}f_{\beta} - \partial_{\beta}f_{\alpha}$.
Similar relations also follow for the vector field $g^\mu$ when considering Eq.~(\ref{eqforg}).

Considering the BC $f_{0} = 0$, we obtain, from Eq.~(\ref{eq_f0}) that 
$\partial^1 f^2 = \partial^2 f^1$
or, using Eqs.~(\ref{eq_f1}) and (\ref{eq_f2}), that

\begin{equation}
\label{eq5_ini_contas}
\partial^{0}\left(\partial^{1} f^{1} + \partial^{2} f^{2}\right)
- \partial^{1} \partial^{1} f^{0} -  \partial^{2} \partial^{2} f^{0} = 0\, .
 \end{equation}

We will make use of a transverse {}Fourier transform for $f_{0}$, similar 
to the one used in Eq.~(\ref{transformadaDeGmenos}),

\begin{equation}
f_{0}(x,y,t) =
\int\frac{d\omega}{2\pi}e^{-i\omega t}
\int\frac{dk}{2\pi} e^{ik y}
\mathcal{F}_{0}(k,\omega,x)\,.
\label{transformadaDe_f0}
\end{equation}

Since we are considering $f_{0} = 0$ at the boundaries, we can write:

\begin{equation}
\label{transformadaDe_f0_atNull_x}
f_{0}(x,y,t)\Bigr|_{x=0\,{\rm and}\, x=a} =
\int\frac{d\omega}{2\pi}e^{-i\omega t}
\int\frac{dk}{2\pi} e^{ik y} 
\mathcal{F}_{0}(k,\omega,x)\Bigr|_{x=0\,{\rm and}\, x=a} = 0\, .
\end{equation}

Since Eq.~(\ref{transformadaDe_f0_atNull_x}) must be valid for all $y$ and $t$,
we conclude that $\mathcal{F}_{0}(k,\omega,x) = 0$ at $x=0$ and $x=a$. 
Thus, we can write

\begin{equation}
\label{derivada2De_f0_atNull_x}
\partial_{2}f_{0}(x,y,t)\Bigr|_{x=0\,{\rm and}\, x=a} =
\int\frac{d\omega}{2\pi}e^{-i\omega t}
\int\frac{dk}{2\pi} e^{ik y} 
i k \mathcal{F}_{0}(k,\omega,x)\Bigr|_{x=0\,{\rm and}\, x=a} = 0\, .
\end{equation}
The condition above has a simple geometric interpretation:  $f_0 (x) = 0$ for all points 
$(0, y)$ and $(a, y)$. Therefore, at $x = 0$ and at $x = a$ the variation of 
$f_0(x, y, t)$ 
with respect to $y$ ($\partial f/\partial y$) is null. 
In a similar way we can conclude that (the following expressions
are to be assumed to be  implicitly valid always at the boundaries,
unless specified otherwise)

\begin{eqnarray}
\label{partial2_g0}
\partial_{2}g_{0} & = & 0\, ,
\\
\label{partial0_partial0_f0}
\partial_{0}\partial_{0}f_{0} & = & 0\, ,
\\
\label{partial2_partial2_f0}
\partial_{2}\partial_{2}f_{0} & = & 0\, .
\end{eqnarray}
{}From Eqs.~(\ref{Afg}), (\ref{derivada2De_f0_atNull_x}) and  (\ref{partial2_g0}), we 
can conclude that the imposition of the BC $F_0 = 0$ is
equivalent to the BC $\partial_{2}A_{0} = 0$. Analogously, we can obtain that
$\partial_{0}A_{0} = 0$.

Also, from Eqs.~(\ref{eq5_ini_contas}) and (\ref{partial2_partial2_f0}), we can write:

\begin{equation}
\label{eq5_contas}
\partial^{0}\left(\partial^{1} f^{1} + \partial^{2} f^{2}\right)
- \partial^{1} \partial^{1} f^{0}= 0 \, .
 \end{equation}

Using Eq.~(\ref{Bianchi}), we can rewrite Eq.~(\ref{eq5_contas}) as
$\partial^{0}\partial_{0} f^{0} - \partial^{1}\partial^{1} f^{0} = 0$.
Thus, using Eq.~(\ref{partial0_partial0_f0}), we can conclude that

\begin{equation}
\label{eq7_contas}
\partial^{1}\partial^{1} f^{0} = 0 \, .
\end{equation}

Making analogous considerations as the ones that lead to  
Eqs.~(\ref{derivada2De_f0_atNull_x}), (\ref{partial0_partial0_f0}) 
and (\ref{partial2_partial2_f0}), we can conclude
from Eq.~(\ref{eq7_contas}) that

\begin{equation}
\label{novaeq7_contas}
\partial_0\partial^{1}\partial^{1} f^{0} = 0 \, .
\end{equation}

Using now Eqs.~(\ref{eq_f0}) and (\ref{eq10_contas}), we 
can write $m_{+}^2 f^{1}  = \partial_{0}f^{10} - m_{+}\partial_{2}f_{0}$.
But since $\partial_{2}f_{0} = 0$, we obtain that

\begin{equation}
\label{eqm2}
m_{+}^2 \, f^{1} = \partial_{0}f^{10} \ \ \ \ \Rightarrow \ \ \ \ 
m_{+}^2\partial^{1}f^{1} = \partial_{0}\partial^1 \partial^1 f^0
- \partial_{0}\partial^0 \partial^1 f^1 \, .
\end{equation}

Using Eq.~(\ref{novaeq7_contas}), we conclude, from Eq.~(\ref{eqm2}), 
that

\begin{equation}
\label{eq11contas_ini}
m_{+}^2\partial_{1}f^{1} = - \partial_{0}\partial^0 \partial_1 f^1 \, .
\end{equation}

We can now also use a transverse {}Fourier transform for ${f^{1}}^{\prime} = 
\partial_{1}f^{1}$, to write

\begin{equation}
{f^{1}}^{\prime}(x,y,t) =
\int\frac{d\omega}{2\pi}\, e^{-i\omega t}
\int\frac{dk}{2\pi}\, e^{ik y}\, 
{\mathcal{F}^1}^{\prime}(k,\omega,x)\,.
\label{transformadaDe_f1_linha}
\end{equation}

Using Eqs.~(\ref{eq11contas_ini}) and (\ref{transformadaDe_f1_linha}), we 
conclude that

\begin{equation}
\label{eq11contas}
m_{+}^2\, {\mathcal{F}^1}^{\prime} = \omega^2\,  {\mathcal{F}^1}^{\prime} \, .
\end{equation}

Since Eq.~(\ref{eq11contas}) must be valid for all $\omega$, we conclude that 
${\mathcal{F}^1}^{\prime} = 0$. Thus, from Eq.~(\ref{transformadaDe_f1_linha}),
we obtain that ${f^{1}}^{\prime} = \partial_{1}f^{1} = 0$.
We can also draw analogous conclusions as applied for the field $g_\mu$.
Thus, we can conclude that the imposition of the BC $F_{0} = 0$ 
(which is here seen in terms of the equivalent strong BC imposed
on the fields $g$ and $f$, i.e., $\tilde{f}_{0} = \tilde{g}_{0} = 0$ and 
$f_{0} = g_{0} = 0$) is equivalent to the BC $\partial_{1}F^{1} = 0$.
Therefore, the same Casimir force should be obtained in the cases of these
two BC. 

We can collect all the results found up to now, to study the behavior of 
$F_{1}$. {}First, using the BC $\partial_{2}A_{0} = 0$
in the definition (\ref{FmuDoA}), we obtain 

\begin{equation}
\label{F1_atBoundaries}
F_{1} = -\partial^0 A^2\, .
\end{equation}
Also, from Eq.~(\ref{FmuDoA}), we deduce that $\partial_{\mu} F^{\mu} = 0$.
But since $\partial_{1}F^{1} = 0$ (at the boundaries), we obtain

\begin{equation}
\label{derivativesOfF0_F2}
\partial_0 F^0 + \partial_2 F^2 = 0\, .
\end{equation}

We are considering the BC $F_{0} = 0$. Then, analogously to what 
we have done in Eqs.~(\ref{transformadaDe_f0_atNull_x}) and 
(\ref{derivada2De_f0_atNull_x}), we can conclude that
(recalling that the relations below are meant to be valid at the boundaries)
$\partial_0 F^0 = 0$. Then, from 
Eq.~(\ref{derivativesOfF0_F2}), we can write that 
$\partial_2 F^2 = 0$. Using again the reasoning 
that lead us from Eq.~(\ref{transformadaDe_f0_atNull_x}) to 
Eq.~(\ref{derivada2De_f0_atNull_x}), we obtain that 
$F^2 = \epsilon^{2 \nu \gamma}\partial_{\nu} A_{\gamma} = 0$.
Using the relation (\ref{Afg}), we conclude that 
$\epsilon^{2 \nu \gamma}\partial_{\nu} g_{\gamma} = 0$
(analogously to $f_{\gamma}$). We can use then the self-duality of
$g_{\gamma}$ (represented by Eq.~(\ref{eqMotion_g}), with $J_{(-)\mu} = 0$)
to obtain $g_{2} = 0$  (analogously to $f_{2}$).

Hence we conclude that $A_{2} = 0$ is also a BC for our model.
Analogously to what we have done above (Eqs.~(\ref{transformadaDe_f0_atNull_x}) and 
(\ref{derivada2De_f0_atNull_x})), we conclude then that
$\partial^0 A_{2} = 0$ and, hence, using
Eq.~(\ref{F1_atBoundaries}), we obtain an equivalent BC: $F_{1} = 0$.

Summarizing, we can conclude that the BC $F_{1} = 0$, $F_{2} = 0$, $\partial_1 F^{1} = 0$ 
and $\partial_2 F^2 = 0$ are all equivalent to the BC $F_{0} = 0$. Therefore, the same 
Casimir force is expected to be obtained for all these cases. Here, we have made explicit 
calculations for the BC $F_{0} = 0$ and $F_{1} = 0$, confirming that the results
obtained are the same in both cases. We note that the particular case for the
Neumann  BC $\partial_{1}F^{1} = 0$ was studied in Ref.~\cite{felipeRudnei}, where it was shown to
also lead to the same result for the
Casimir force, Eq.~(\ref{ffinal})\footnote{It should be noticed that in Ref.~\cite{felipeRudnei}
there is a misprint in the expression for the masses $m_1$ and $m_2$ considered there
by a factor two. With this correction, those two masses considered in that reference 
just correspond to $m_\pm$ considered
here. This in turn corresponds to a correction in Sec. IV of that reference, where
the CS parameter considered there should be replaced by $2 \Theta$ instead.}.

\section{Suppression of the Casimir force in the presence of vortex particle-like excitations}
\label{sec6}

As shown in the previous sections, the Casimir force for the cases of PC ($F_1=0$), 
MP ($F_0=0$) and also Neumann ($\partial_1 F^1 =0$) BC all lead
to the same result,

\begin{equation}
\label{casimirfinal}
f =-\frac{1}{16\pi a^3}\left[\int_{2am_{-}}^{\infty}dz \frac{z^2}{e^z-1}
+\int_{2am_{+}}^{\infty}dz \frac{z^2}{e^z-1}\right].
\end{equation}

Note that Eq.~(\ref{casimirfinal}) is of the form of a
second Debye function~\cite{abra},

\begin{equation}
\int_{b}^{\infty}dz\  \frac{z^{n}}{ e^z - 1} = \sum_{k=1}^{\infty}
e^{-k b} \left(\frac{b^n}{k} + n \frac{b^{n-1}}{k^2} +   n(n-1)
\frac{b^{n-2}}{k^3} + \ldots + \frac{ n!}{k^{n+1}} \right)\,,
\label{debye}
\end{equation}
indicating that the Casimir force for both cases decays exponentially
with $a m_\pm$.

Specific limits for $a m_{\pm}$, like for small or large values, can
be  easily derived using directly the expression (\ref{ffinal}) or 
from (\ref{debye}). These results can also be readily
expressed in terms of the Proca and Chern-Simons masses, $m$ and
$\mu$, respectively, using  Eq.~(\ref{mMenosMais}), or also from
Eqs.~(\ref{massm}) and (\ref{massmu}), relating these masses to the
original parameters of the effective particle-vortex dual Lagrangian
density model.

By expressing $m_\pm$ in terms of the the original parameters of the
particle-vortex dual Lagrangian density model, i.e., in terms of  the
vacuum expectation values for the Higgs field, $\rho_0$, for the
vortex field,  $\psi_0$, and the CS parameter $\Theta$,  we have that

\begin{eqnarray}
m_\pm = \frac{e^2 \rho_0^2}{2\Theta} \left[ \sqrt{1+ \left(8 \pi
    \frac{\Theta\psi_0}{e^2 \rho_0}\right)^2} \mp 1\right]\,.
\end{eqnarray}

As it was shown in Ref.~\cite{dual}, vortices are energetically
favored to condense for values of the CS parameter below a critical
value $\Theta_c \approx (e^2/\pi)\ln 6 \simeq 0.57 e^2$. {}For $\Theta
< \Theta_c$ the vortex condensate can be written as $\psi_0^2 \approx
(e^2 \rho_0^2/\Theta) \sqrt{6-\exp(\pi \Theta/e^2)}$. The condensed
vortex  phase can be interpreted as being equivalent to the Shubnikov
phase for  type-II superconductors  in the presence of a magnetic
field~\cite{tink}, with a Ginzburg-Landau parameter   $\kappa \equiv e
\rho_0/\Theta > 1/\sqrt{2}$. In the analysis that follows,  we will
remain within parameter values satisfying these conditions. 

\begin{figure}[htb]
  \vspace{0.75cm}
  \epsfig{figure=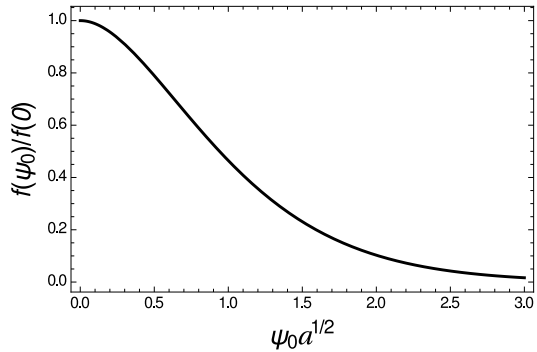,angle=0,width=7cm}
\caption[]{\label{fig1} The (normalized) Casimir force as a function of the vortex
  condensate $\psi_0$.  The following representative values of parameters
  were used:  $\Theta/e^2=0.1$ and $\rho_0 a^{1/2} =1$.}
\end{figure}
In {}Fig.~\ref{fig1} we show the result for the Casimir force
Eq.~(\ref{casimirfinal}), as a function of the vortex
condensate $\psi_0$, normalized by the Casimir force in the absence
of a vortex condensate, $f(\psi_0=0)$.
The result shows that the  Casimir force can become strongly suppressed in the
presence of vortex  matter as compared to the absence of it.
This suppression of the Casimir
force can be interpreted as a result of the repelling force between
vortices, analogously to what happens in the phenomenology of type-II
superconductors, when in the Shubnikov phase~\cite{tink}, which
opposes  the attractive Casimir force.  

\section{Conclusions}
\label{sec7}

In this work, we have analyzed the Casimir force for the MPCS model. As explained
in Sec.~\ref{sec2}, this model can be interpreted as an effective (dual) model describing
vortex excitations for a CSH model.  We have obtained the Casimir
force for the cases of perfect conductor and perfect magnetically
permeable BC. This has been possible by mapping the MPCS model into a
doublet consisting of a self-dual and an anti self-dual PCS
models. The result obtained for the Casimir force was found to be
the same for both cases of BC used, which also
agrees with the case of considering the Neumann BC, which was derived 
previously in Ref.~\cite{felipeRudnei}.
The reason for these results to be the same has been explained to be
a consequence of the symmetry and constraints satisfied by these models
involving a CS term. These results have also been confirmed by using
the mapping of the MPCS model in a doublet of MCS models. The derivation
using these two independent mappings also helps to
show that the result obtained for the Casimir force (for the type
of BC considered here) is not some particular consequence of the mapping used.
Thus, our results also highlight a symmetry found when we consider various types of BC in
the computation of the Casimir  effect.

Even though it can be argued that the model we have studied here, which can
be associated with the vacuum state of a system of vortex excitations
in a plane, is mostly of theoretical interest and might be far from
describing  real physical systems of interest, our results are,
however, indicative of a behavior that can manifest in these systems.
As such, our results might be of relevance for the next generation of
experiments involving the Casimir effect~\cite{Allocca:2012kw}, or
those involving, for example,  vortex-based superconducting
detectors~\cite{AMKadin,AlexeiDSemenovaEtAl_1}.  Usually, such systems
involve nanometer scales, in which the Casimir force turns out to be
relevant, and possibly also altering the microscopic parameters of the
detectors~\cite{J.OfLowTemp.Phys.151}.  Our results can also be of
relevance when devising materials based on superconducting films to
work as possible suppressors of the Casimir force, such as in those
laboratory experiments that require  performing extremely careful
force measurements near surfaces. This might be the case of the
searches  for possible deviations of the Newtonian gravity. 

The study performed here for the MPCS model also has its own merits,
independent of its connection to a vortex model. The MPCS model
constitutes of massive gauge particles, with mass terms that have both
topological and non-topological  origins. Also, the Maxwell-Proca and
the MCS models can be seen as particular cases of the MPCS model. So,
we expect that a better comprehension of the roles of the mass terms,
be them either of topological or non-topological origin, in the derivation of
the Casimir force might eventually provide arguments in favor of one or the other,
when using these models with the objective of understanding some of
the properties of real planar systems with massive excitations. This
also  includes, of course, deriving the Casimir force under different
BC,  as we have studied in this work. 


\acknowledgments

R.O.R. is partially supported by Conselho Nacional de Desenvolvimento
Cient\'{\i}fico e Tecnol\'{o}gico (CNPq-Brazil) and by {}Funda{\c
  {c}}{\~{a}}o de Amparo {\`{a}} Pesquisa do Estado do Rio de Janeiro
(FAPERJ). V. S. A. is partially supported by Coordena{\c {c}}{\~{a}}o
de Aperfei{\c {c}}oamento de Pessoal de N\'{\i}vel Superior
(CAPES-Brazil). 
R.F.O. and C.R.M.S are partially supported by Conselho Nacional de Desenvolvimento
Cient\'{\i}fico e Tecnol\'{o}gico (CNPq-Brazil). 
J.F.M.N. thanks J.A. Helay\"{e}l Neto, S. Perez, M. C. de Lima and
D. T. Alves for very fruitful commentaries.
We are also grateful to the anonymous referee for the many helpful comments,
that allowed us to improve the manuscript from its first version.


\end{document}